\documentclass[preprint,prb]{revtex4}

\usepackage{graphicx,subfigure}		
\usepackage{amssymb,amsmath}		

\newcommand{\fullwidth}		{5.0in}
\newcommand{\halfwidth}		{2.8in}

\newcommand{\BB}[1]		{\mbox{\boldmath${#1}$}}
\newcommand{\modulo}[1]	{{\left \vert {#1} \right \vert}}
\newcommand{\fracn}[2] 		{{\textstyle\frac{#1}{#2}}}

\newcommand{\Vd}		{{\BB{d}}}
\newcommand{\Ve}		{{\BB{e}}}

\newcommand{\Vi}		{{\BB{i}}}
\newcommand{\Vj}		{{\BB{j}}}
\newcommand{\Vk}		{{\BB{k}}}
\newcommand{\Vl}		{{\BB{l}}}

\newcommand{\Vp}		{{\BB{p}}}
\newcommand{\Vq}		{{\BB{q}}}
\newcommand{\Vr}		{{\BB{r}}}

\newcommand{\Vu}		{{\BB{u}}}
\newcommand{\Vv}		{{\BB{v}}}

\newcommand{\VF}		{{\BB{F}}}

\newcommand{\VP}		{{\BB{P}}}
\newcommand{\VQ}		{{\BB{Q}}}

\newcommand{\VY}		{{\BB{Y}}}

\newcommand{\Vone}		{{\BB{1}}}

\newcommand{\Vomega}		{{\BB{\omega}}}
\newcommand{\Vphi}		{{\BB{\phi}}}

\newcommand{\VGamma}		{{\BB{\Gamma}}}
\newcommand{\VOmega}		{{\BB{\Omega}}}

\newcommand{\Ds}		{\partial_s}
\newcommand{\Dt}		{\partial_t}

\newcommand{\vs}		{{\it vs. }}
\newcommand{\cf}		{{\it c.f. }}

\begin{document}
\preprint{}

\title{A symplectic integration method for elastic filaments}
\author{Anthony J. C. Ladd} \email{tladd@che.ufl.edu}
  \homepage{http://ladd.che.ufl.edu/} \author{Gaurav Misra}
\email{gmisra@che.ufl.edu} \affiliation{ Department of Chemical
Engineering, University of Florida, Gainesville, FL 32611}
\date{\today}

\begin{abstract}
A new method is proposed for integrating the equations of motion of an
elastic filament. In the standard finite-difference and finite-element
formulations the continuum equations of motion are discretized in
space and time, but it is then difficult to ensure that the
Hamiltonian structure of the exact equations is preserved. Here we
discretize the Hamiltonian itself, expressed as a line integral over
the contour of the filament. This discrete representation of the
continuum filament can then be integrated by one of the explicit
symplectic integrators frequently used in molecular dynamics. The
model systematically approximates the continuum partial differential
equations, but has the same level of computational complexity as
molecular dynamics and is constraint free. Numerical tests show that
the algorithm is much more stable than a finite-difference formulation
and can be used for high aspect ratio filaments, such as actin.
\end{abstract}
\maketitle

\section{Introduction}\label{sec:intro}
Elastic rods are a ubiquitous model of semi-flexible
biopolymers such as
DNA,\cite{Marko1994a,Marko1995,Swigon1998,Balaeff1999,Coleman2000,Tobias2000a,Coleman2003,Rossetto2003}
actin,\cite{Isambert1995,Maggs1998,Gardel2004,DiDonna2007} and
microtubules.\cite{Brangwynne2006} They can also be found in a
diverse range of applications including catheter
navigation,\cite{Lawton1999} undersea
cables,\cite{Goyal2008} and organismal
biology.\cite{Goriely2006} In biophysics, the worm-like chain
(WLC) model\cite{Marko1994,Marko1995} underpins many
theoretical\cite{Levine2004a,Head2005,Storm2005,DiDonna2006,DiDonna2007,Das2007,Mizuno2007,MacKintosh2008}
and numerical\cite{Everaers1999,Wada2006,Wada2007a,Llopis2007} studies of
semiflexible polymers. The WLC model is a linearization of the
classical Kirchoff rod model,\cite{Love1944,Landau1959} which
is itself a limiting case where the product of the local
curvature and filament thickness is everywhere
small.\cite{Coleman1993} In this limit the shear and
extensional strains are negligible but the constraint forces
generated by them are not. In this paper we consider a
generalization of the Kirchoff
model,\cite{Romero2002,Bishop2004} where the shear and
extensional strains are explicitly accounted for by an elastic
constitutive model, eliminating the need for constraint forces
at the cost of an additional time scale; such models are
frequently referred to as ``geometrically exact'' in the
finite-element literature.\cite{Romero2002,Bishop2004}

The dynamics of Kirchoff or geometrically exact (GE) filaments is
typically determined by finite-element or finite-difference
approximations, but the stiffness of the numerical system has proved
to be a difficult and long-standing
problem.\cite{Simo1988,Simo1995} Significant progress has been made
by developing implicit methods that exactly satisfy the constraints of
momentum and energy conservation,\cite{Romero2002,Goyal2005} yet even
here artificial dissipation is often needed for long-term
stability.\cite{Armero2003} On the other hand, in {\em discrete}
dynamical systems it is known that symplectic integration methods give
superior long-term stability in comparison with either high-order
explicit or implicit integration methods;\cite{Dullweber1997} the
most common symplectic integrator is the Verlet
algorithm.\cite{Verlet1967} Symplectic integrators generate a
sequence of canonical transformations, which do not exactly conserve
energy but do preserve the density of points in the phase space, along
with the Poincar\'e invariants. In recent years symplectic integrators
have been developed for both linear and angular
motions.\cite{Dullweber1997,Miller2002,Zon2008} The objective of this
paper is to explore a symplectic integration method for geometrically
exact filament models. This requires both a Hamiltonian approximation
to the partial differential equations describing the filament
dynamics, and a symplectic integrator.

The proposed algorithm is based on a discretization of the Hamiltonian
line integral of an elastic filament, including shear and extensional
degrees of freedom. Since the nodal forces and torques follow from an
{\em exact differentiation} of a potential function, the equations of
motion are guaranteed to be Hamiltonian, although the potential
function itself is only an approximation to the continuum limit. This
is in contrast to finite-element methods, where the continuum
equations of motion are discretized in space; in this case the
Hamiltonian structure is not preserved, even if the total energy is
conserved.\cite{Romero2002} In fact, it can be shown that for any
approximate solution it is not possible to maintain both the
symplectic structure and exact energy conservation
simultaneously.\cite{Zhong1988}

An outline of the paper is as follows. In Sec.~\ref{sec:models} we
describe different models of elastic filaments--GE, Kirchoff, WLC--and
indicate how they are related. Next (Sec.~\ref{sec:DEOM}), we derive a
simple finite-difference approximation of the equations of motion of a
GE filament model, as a basis for comparison with the Hamiltonian
formulation presented in Sec.~\ref{sec:H}. We note that the
Hamiltonian approach has only been followed
occasionally,\cite{Dichmann1996} and in that case for the Kirchoff
rod model.  We will argue (Sec.~\ref{sec:example}) that the absence of
geometric constraints in the GE model offers computational advantages
over the Kirchoff model when there are excluded volume interactions
between the segments. We replace the usual implicit time
integration\cite{Dichmann1996,Romero2002} with an explicit operator
splitting method,\cite{Miller2002} which eliminates the repeated
force evaluations of an implicit method. The numerical scheme is
stable and energy conserving even for large deformations; we
illustrate this by numerical example in Sec.~\ref{sec:example}. Our
conclusions and future outlook are in Sec.~\ref{sec:conc}.

\section{Elastic filament models}\label{sec:models}

\begin{figure}
\center \includegraphics[width=\fullwidth]{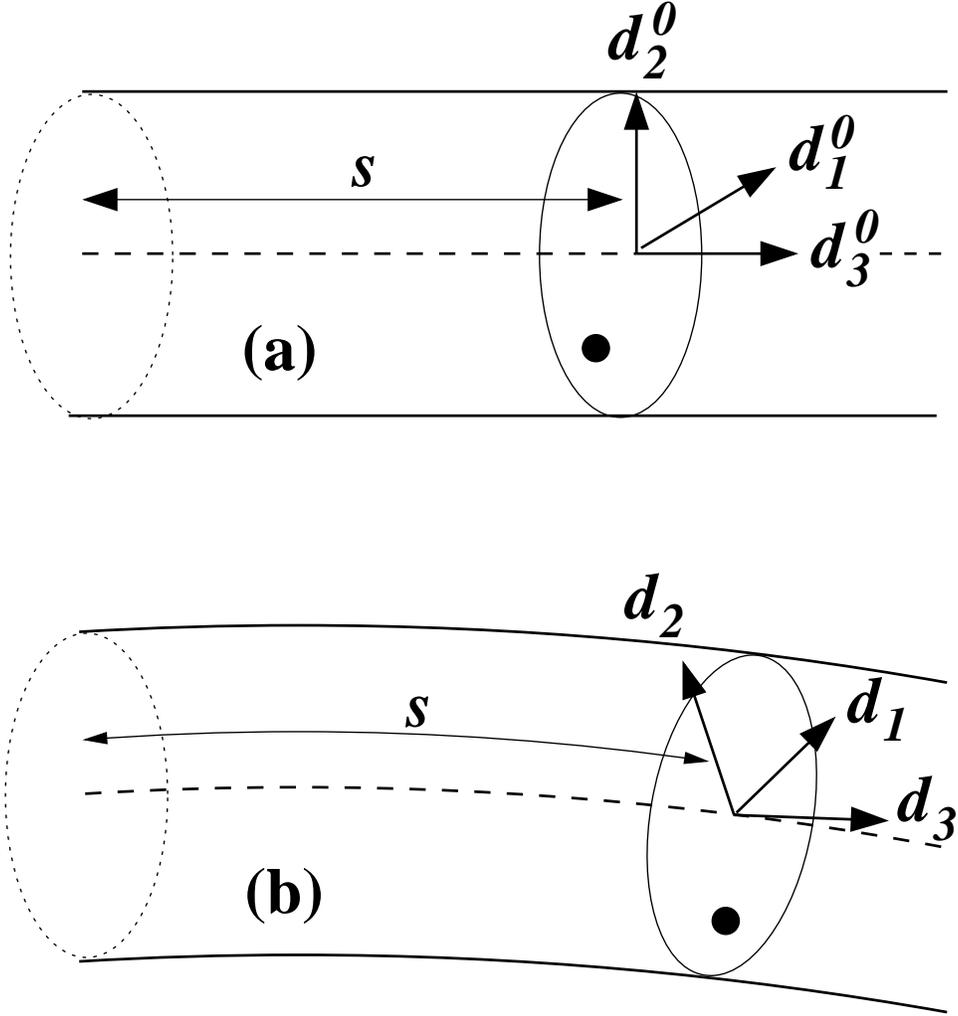}
\caption{An elastic filament in the unstrained (reference) state (a)
and after deformation (b). In the reference state, the material plane,
shown by the solid ellipse, is aligned with its normal parallel to the
tangent to the centerline (dashed line). The local director basis of
the reference state, $\Vd_i^0(s)$, and the deformed state, $\Vd_i(s)$,
are also shown. A material point (solid black circle) moves with the
translation and rotation of the local coordinate system; in this case
extension, shear, bend, and twist can all be seen.}
\label{fig:rodmodel}
\end{figure}

The classical Kirchoff theory of elastic rods has been elegantly and
concisely described in the ``{\em Theory of Elasticity}'' by Landau
and Lifshitz,\cite{Landau1959} and the seminal book by
Love.\cite{Love1944} More rigorous derivations of the equations of
motion are available in the
literature.\cite{Green1966,Coleman1993} Here we summarize the key
concepts and establish the notation to be used later in the paper. An
elastic filament (or thin rod) is described by the coordinates of its
centerline $\Vr(s)$ and a set of orthonormal directors $\Vd_1(s)$,
$\Vd_2(s)$, $\Vd_3(s)$. The directors establish the orientation of a
cross section or material plane at the location $s$, where $s$ is a
parametric coordinate defining the position of each point along the
centerline. In the undeformed filament, $s$ is the contour length from
the origin. We will choose a body-fixed coordinate system such that
$\Vd_1$ and $\Vd_2$ point along the principal axes of inertia of the
cross section and therefore $\Vd_3 = \Vd_1 \times \Vd_2$ is normal to
the material plane; the coordinate system is illustrated in
Fig~\ref{fig:rodmodel}. If the rod has a circular cross section then
the initial choice of $\Vd_1$ and $\Vd_2$ contains an arbitrary
rotation about $\Vd_3$. In contrast with the Kirchoff theory, we will
not assume that $\Vd_3$ is constrained to be parallel to the tangent
vector $\Ds \Vr$ (Fig~\ref{fig:rodmodel}b).

The key assumption of thin-rod elasticity is that there is no
deformation within a material plane, only translation and rotation of
that plane. Deformation of an elastic filament is then described by
two one-dimensional strain fields, $\VGamma(s)$ and $\VOmega(s)$,
describing the rate of change of the centerline position and director
vectors along the filament\cite{Romero2002,Bishop2004}
\begin{equation}
\begin{array}{ccc}
  \Gamma_1 = \Vd_1 \cdot (\Ds \Vr) & ~~~ & \Omega_1 = \Vd_3 \cdot (\Ds
  \Vd_2)  = - \Vd_2 \cdot (\Ds \Vd_3)  \\ \Gamma_2 = \Vd_2 \cdot (\Ds
  \Vr) & ~~~ & \Omega_2 = \Vd_1 \cdot (\Ds \Vd_3) = -  \Vd_3 \cdot
  (\Ds \Vd_1) \label{eq:strains} \\ \Gamma_3 = \Vd_3 \cdot (\Ds \Vr) &
  ~~~ & \Omega_3 = \Vd_2 \cdot (\Ds \Vd_1) = -  \Vd_1 \cdot (\Ds
  \Vd_2).
\end{array}
\end{equation}
A thin segment of the filament can be subjected to six different
deformations. $\Gamma_1$ and $\Gamma_2$ describe transverse motions of
a material plane with respect to the normal vector ($\Vd_3$), which
causes shearing of the segment, while $\Gamma_3$ describes extension
or compression of the segment. Bending of the segment about its
principal axes is described by $\Omega_1$ and $\Omega_2$, and twisting
of the segment by $\Omega_3$. Uniform deformation corresponds to
constant values of $\VGamma$ and $\VOmega$; for example, in a straight
rod $\VGamma = [0,0,1]$ and $\VOmega = [0,0,0]$. More interestingly, a
helical rod can be described by a constant bend and twist, $\VGamma =
[0,0,1]$, $\VOmega = [R \kappa^2,0,P\kappa^2]$, where $R$ is the
radius of the helix, $2 \pi P$ is the pitch, and the combined
curvature due to bend and twist, $\kappa = (P^2+R^2)^{-1/2}$. The
choice of signs define a right-handed helix, $\Vr(s) = [R \cos (\kappa
s), R \sin (\kappa s), P\kappa s]$, with basis vectors
\begin{equation}
\begin{array}{l}
 \Vd_1 = [P\kappa\sin(\kappa s), -P\kappa\cos(\kappa s), R\kappa] \\
 \Vd_2 = [\cos(\kappa s), \sin(\kappa s), 0] \label{eq:dhelix} \\
 \Vd_3 = [-R\kappa\sin(\kappa s), R\kappa\cos(\kappa s), P\kappa].
 \end{array}
\end{equation}

The stresses in the rod are assumed to be linear in the deviations in
the strain fields, $\Delta \Gamma_i = \Gamma_i - \Gamma_i^0$ and
$\Delta \Omega_i = \Omega_i - \Omega_i^0$, from the reference (stress
free) configuration $\VGamma^0$, $\VOmega^0$. It is convenient to
define the strains in the body-fixed coordinate system, since the
elastic constant matrix is then diagonal. The force $F_i^\Gamma$ and
couple $F_i^\Omega$ on each material plane
are\cite{Landau1959,Love1944}
\begin{equation}\label{eq:FC}
F_i^\Gamma = C_i^\Gamma \Delta \Gamma_i, ~~~ F_i^\Omega =  C_i^\Omega
\Delta \Omega_i,
\end{equation}
where the elastic constants for each deformation are, in principle,
independent. In the GE model, the strain energy density $U(s)$
contains contributions from shear and extension, in addition to the
usual bend and twist of the Kirchoff model,
\begin{equation}\label{eq:U}
 U = U^\Gamma + U^\Omega = \frac{1}{2} \sum_{i=1}^3 \left(C_i^\Gamma
 \Delta\Gamma_i^2    + C_i^\Omega \Delta\Omega_i^2 \right).
\end{equation}

For an isotropic material, the elastic moduli for shear
($C_{1,2}^\Gamma$), extension ($C_3^\Gamma$), bend ($C_{1,2}^\Omega$),
and twist ($C_3^\Omega$) are given by:
\begin{equation}\label{C}
\begin{array}{ccc}
  C_1^\Gamma = G A & ~~~ & C_1^\Omega = Y I_1 \\ C_2^\Gamma = G A &
  ~~~ & C_2^\Omega = Y I_2 \\ C_3^\Gamma = Y A & ~~~ & C_3^\Omega = G
  I_3,
\end{array}
\end{equation}
where $G$ is the shear modulus, $Y$ is Young's modulus, $A$ is the
area of the cross-section and $I_1$ and $I_2$ are its principle
moments of inertia. For rods with a circular cross section, $I_3 = I_1
+ I_2$, but in the general case there is an additional contribution
from the warping of the cross section,\cite{Landau1959} so that $I_3$
is then distinct from $I_1 + I_2$. The elastic coefficients can also
be determined empirically, without reference to any particular
constitutive law.

The velocity and angular velocity of the segment are defined in an
analogous fashion to the strain fields in Eq.~\ref{eq:strains},
\begin{equation}
\begin{array}{ccc}
  v_1 = \Vd_1 \cdot (\Dt \Vr) & ~~~ & \omega_1 = \Vd_3 \cdot (\Dt
  \Vd_2)  = - \Vd_2 \cdot (\Dt \Vd_3)  \\ v_2 = \Vd_2 \cdot (\Dt \Vr)
  & ~~~ & \omega_2 = \Vd_1 \cdot (\Dt \Vd_3) = -  \Vd_3 \cdot (\Dt
  \Vd_1) \label{eq:vels} \\ v_3 = \Vd_3 \cdot (\Dt \Vr) & ~~~ &
  \omega_3 = \Vd_2 \cdot (\Dt \Vd_1) = -  \Vd_1 \cdot (\Dt \Vd_2).
\end{array}
\end{equation}
The kinetic energy density of the filament is
then\cite{Landau1959,Love1944}
\begin{equation}\label{eq:T}
 T = T^\Gamma + T^\Omega = \frac{1}{2} \sum_{i=1}^3  \left(M_i^\Gamma
v_i^2 + M_i^\Omega \omega_i^2 \right),
\end{equation}
where the generalized mass densities associated with shear
($M_1^\Gamma, M_2^\Gamma$), extension ($M_3^\Gamma$), bend
($M_1^\Omega, M_2^\Omega$) and twist ($M_3^\Omega$), are
\begin{equation}\label{eq:M}
\begin{array}{ccc}
  M_i^\Gamma  = \rho A, & ~~~ & M_i^\Omega = \rho I_i,
\end{array}
\end{equation}
and $\rho$ is the mass density of the filament.

Equations of motion for the filament can be derived from the balance
of linear and angular momenta in a thin segment bounded by the planes
$s$ and $s+ds$. The rate of change of the linear momentum of the
segment, $\Vp ds$, is
\begin{equation}
{\dot \Vp} ds = \VF^\Gamma(s+ds)-\VF^\Gamma(s),
\end{equation} 
where $\Vp =  \sum_{i=1}^3 M^\Gamma v_i \Vd_i$ is the linear momentum
density (per unit length).  The forces on the two planes must be
differenced in a common coordinate frame, which we take as the
space-fixed frame. The balance of angular momentum in the segment $\Vl
ds$ involves both couples and moments of the force,
\begin{equation}
{\dot \Vl} ds = \VF^\Omega(s+ds)-\VF^\Omega(s)  + \Vr(s+ds) \times
\VF^\Gamma(s+ds) -\Vr(s) \times \VF^\Gamma(s),
\end{equation}
where $\Vl = \sum_{i=1}^3 M_i^\Omega \omega_i \Vd_i$ is the linear
angular momentum density. Thus the equations of motion of a GE
filament are
\begin{eqnarray}
{\dot \Vp} &=& \Ds \VF^\Gamma\label{eq:EOMp}, \\ {\dot \Vl}&=&  \Ds
\VF^\Omega + \Vr^\prime \times \VF^\Gamma \label{eq:EOMl},
\end{eqnarray}
where $\Vr^\prime = \Ds \Vr$ indicates a spatial derivative along the
filament. A finite-difference approximation to these equations is
described in Sec.~\ref{sec:DEOM}.

Equations~\ref{eq:EOMp}--\ref{eq:EOMl} describe the dynamics of the GE
rod model.\cite{Romero2002,Bishop2004} The difference with the
Kirchoff theory is that, here, the force on a material plane,
$F_i^\Gamma$, is given by a constitutive equation, Eq.~\eqref{eq:FC},
based on the deflection and extension of the local tangent vector
relative to the material plane, Eq~\eqref{eq:strains}. In the Kirchoff
model the tangent vector is constrained to remain parallel to $\Vd_3$
(unshearable) and of unit length (inextensible), or in other words
$\Delta \Gamma_i = 0$ and $\Vr^\prime = \Vd_3$. As a result,
neighboring segments can only rotate with respect to one another,
leading to a compatibility condition,\cite{Goyal2005}
\begin{equation}
\Vv^\prime = \Vomega \times \Vr^\prime = {\dot \Vd}_3,
\label{eq:compat}
\end{equation}
where the last equality follows from the kinematic
conditions, ${\dot \Vd}_i = \Vomega \times
\Vd_i$.\cite{Landau1959,Green1966}  Differentiating Eq.~\eqref{eq:EOMp} with respect to $s$ gives
an equation for the constraint force satisfying the compatibility
equation,
\begin{equation}\label{eq:constraint12}
\Ds^2 \VF^\Gamma = M^\Gamma {\ddot \Vd}_3,
\end{equation}
where ${\ddot \Vd}_3 = {\dot \Vomega} \times \Vd_3 + \Vomega \times
(\Vomega \times \Vd_3)$.\cite{Lawton1999,Gueron2001a,Goyal2005} A
simpler, but approximate solution is to neglect the angular momentum
perpendicular to the tangent vector,\cite{Klapper1996,Hou1998} and
determine the shear forces, $\VF^{\Gamma,\perp}$, directly from the
cross product of Eq.~\eqref{eq:EOMl} with $\Vd_3$,
\begin{equation}
\Vd_3 \times \Ds \VF^\Omega = \left( \Vone - \Vd_3\Vd_3 \right) \cdot
\VF^\Gamma = \VF^{\Gamma,\perp}. \label{eq:constraint2}
\end{equation} 
The force along $\Vd_3$ is determined from the inextensibility
condition,\cite{Tornberg2004}
\begin{equation} \label{eq:constraint}
\Ds \Vr \cdot \Ds \Vr = 1.
\end{equation}
The Kirchoff model has the computational advantage that the shear and
extensional modes are frozen by the constraints, so that a larger time
step may be used. On the other hand the numerical integration is
inherently implicit and must be solved iteratively at each time step.

Bending forces can also be determined from the curvature in the
centerline position vector,\cite{Landau1959} $\Vr^\prime \times
\Vr^{\prime\prime}$, rather than from derivatives of the basis
vectors, Eq.~\eqref{eq:strains}. In the case of a weakly bent
filament, the tangent can be assumed to be locally
constant,\cite{Landau1959} and, with an isotropic bending stiffness
$C_1^\Omega = C_2^\Omega = C^\Omega$,
\begin{equation}
\VF^{\Gamma,\perp} = - C^\Omega \left( \Vone -
\Vr^{\prime}\Vr^{\prime}\right) \cdot \Vr^{\prime\prime\prime}.
\end{equation} 
Differentiating once more (again ignoring derivatives of
$\Vr^\prime$), we obtain the equation of motion for the bending of a
WLC,\cite{Everaers1999,Tornberg2004,Wada2006,Llopis2007}
\begin{equation}
M^\Gamma {\ddot \Vr} = - C^\Omega \left( \Vone -
\Vr^{\prime}\Vr^{\prime}\right) \cdot \Vr^{\prime\prime\prime\prime},
\label{eq:WLCEOM}
\end{equation}
although what is really being calculated is the constraint force
needed to resist the shear deformations arising from the compatibility
condition, Eq.~\ref{eq:compat}. In addition, a constraint force is
needed to satisfy the inextensibility condition,
Eq.~\eqref{eq:constraint}. Unfortunately, Eq.~\ref{eq:WLCEOM} is very
stiff, and numerical integration of the partial differential equations
is not straightforward.\cite{Tornberg2004} Most simulations of the
WLC model have therefore discretized the filament into a sequence of
beads interacting via a bending
potential.\cite{Everaers1999,Wada2006,Llopis2007} Although this
sacrifices fidelity to the continuum filament model, the ordinary
differential equations for the bead positions can be integrated using
standard molecular dynamics methods, including constraint forces to
maintain a discrete approximation to Eq.~\eqref{eq:constraint}. In
this paper we derive a discrete Hamiltonian representation of a GE rod
model, along the lines already established for the WLC. Our algorithm
systematically approximates the GE filament model, while maintaining
the simplicity of the WLC approach. We wish to emphasize that the
models described in this work are discrete approximations to
continuous filaments, in which the nodes indicate representative
points along the centerline. This is different from models where the
segments are physical objects with finite length, undergoing
rigid-body motion.\cite{Butler2005,Qi2006}

\section{Discrete equations of motion}\label{sec:DEOM}

We first describe a spatial discretization of the equations of motion
of a GE rod, Eqs.~\ref{eq:EOMp}--\ref{eq:EOMl}. The filament is
divided into $N$ equal segments of length $\Delta s = L/N$, and nodes
are defined at the center of each segment,\cite{Dichmann1996}
\begin{equation}
s_n = \left(n - \fracn{1}{2} \right) \Delta s, ~~~ n = 1, 2, \ldots N.
\end{equation}
The instantaneous state of the filament is then given by the nodal
coordinates $r_\alpha^n$, quaternions $q_a^n$, linear momenta
$p_\alpha^n$, and angular momenta $l_i^n$. We use Greek subscripts,
$\alpha, \beta, \gamma$, to indicate components in the space-fixed
frame, subscripts $i, j, k$, to indicate components in the body-fixed
frame, and the subscripts $a, b, c$, to denote the components of the
quaternion, $q_a = [q_0, q_x, q_y, q_z]$. The Einstein summation
convention is applied to the subscripts $\alpha, \beta, \gamma$ and
$a, b, c$, but not to the indexes $i, j, k$. Thus for example
\begin{equation}
p_\alpha = \sum_{i=1}^3 p_i d_{i\alpha}, ~~~ p_i = d_{i\alpha}
p_\alpha.
\end{equation}

\begin{table}
\caption{Properties of Quaternions (Appendix~\ref{app:A})}
\renewcommand{\thesection}{T\arabic{section}}
\begin{equation*}\tag{T1.1}\label{eq:euler2q}
\begin{array}{c}
	q_0 = \cos \left( \frac{\vartheta}{2} \right) \cos \left(
	\frac{\phi+\psi}{2} \right) \\ q_x = \sin \left(
	\frac{\vartheta}{2} \right)  \cos \left( \frac{\phi-\psi}{2}
	\right) \\ q_y = \sin \left( \frac{\vartheta}{2} \right)  \sin
	\left( \frac{\phi-\psi}{2} \right)  \\ q_z = \cos \left(
	\frac{\vartheta}{2} \right) \sin \left( \frac{\phi+\psi}{2}
	\right)
\end{array}
\end{equation*}
\center{Relation between quaternions and Euler angles $(\phi,
\vartheta, \psi)$\cite{Landau1976,Goldstein1980}}

\begin{equation*}\tag{T1.2}
        \left(
                \begin{array}{c} \Vd_1 \\ \Vd_2 \\ \Vd_3 \end{array}
        \right) = \left(
                \begin{array}{ccc}
                        q_0^2+q_x^2-q_y^2-q_z^2 & 2(q_x q_y+q_0 q_z) &
                        2(q_x q_z-q_0 q_y) \\ 2(q_y q_x-q_0 q_z) &
                        q_0^2-q_x^2+q_y^2-q_z^2 & 2(q_y q_z+q_0 q_x)
                        \\ 2(q_z q_x+q_0 q_y) & 2(q_z q_y-q_0 q_x) &
                        q_0^2-q_x^2-q_y^2+q_z^2
                \end{array}
        \right) . \label{eq:di}
\end{equation*}
\center{Director basis in terms of quaternions}

\begin{equation*}\tag{T1.3}
        \left(
                \begin{array}{c} \Ve_1 \\ \Ve_2 \\ \Ve_3 \end{array}
        \right) = \left(
                \begin{array}{cccc}
                         -q_x & ~~\,q_0 & ~~\,q_z &  -q_y\\ -q_y &
                         -q_z & ~~\,q_0 & ~~\,q_x\\ -q_z & ~~\,q_y &
                         -q_x & ~~\,q_0
                \end{array}
        \right) .  \label{eq:ei}
\end{equation*}
\center{Body-fixed rotations in a quaternion basis}

\begin{equation*}\tag{T1.4}
     \frac{\partial d_{i\alpha}}{\partial q_a} = \sum_{j,k=1}^3
     2\epsilon_{ijk}d_{j\alpha}e_{ka} + 2q_a d_{i\alpha}.
     \label{eq:ddi}
\end{equation*}
\center{Derivatives of $\Vd$ vectors}

\begin{equation*}\tag{T1.5}
     \frac{\partial e_{ia}}{\partial q_b} = \sum_{j,k=1}^3
     \epsilon_{ijk}e_{ja}e_{kb} + e_{ia}q_b - q_a e_{ib}.
     \label{eq:dei}
\end{equation*}
\center{Derivatives of $\Ve$ vectors}
\label{tb:q}
\end{table}

The quaternion ${\cal Z} = [q_0, \Vq]$ describes a rotation about an
axis parallel to the vector $\Vq = [q_x, q_y, q_z]$ by an angle
$\vartheta = 2\cos^{-1}(q_0)$. The orientation of a body in space can
be specified by the components of ${\cal Z}$, which we denote by
$q_a$. We use quaternions in preference to the director basis vectors
as angular coordinates,\cite{Romero2002} since it reduces the number
of degrees of freedom. Symplectic integration algorithms using
operator splitting exist for both quaternions\cite{Miller2002} and
director vectors.\cite{Dullweber1997} The choice of the body-fixed
angular momenta is guided by the integration
algorithm,\cite{Miller2002} which requires them for the quaternion
update. Key properties of quaternions are summarized in
Table~\ref{tb:q} and derived in Appendix~\ref{app:A}.

An infinitesimal rotation about the body-fixed axes can be written in
terms of variations in the quaternions (see Appendix~\ref{app:A} for
details),
\begin{equation}\label{eq:q2phi}
\delta \phi_i = 2e_{ia} \delta q_a,
\end{equation} 
where the quaternion variation is subject to the normalization
constraint $\delta q_a q_a = 0$. In other words the variation in $q_a$
must be in a three-dimensional space orthogonal to $q_a$. The
quaternion basis vectors $e_i$ (Eq.~\ref{eq:ei}) describe rotations
about a body-fixed axis and are orthogonal to each other and to the
quaternion itself. The factor of 2 arises because it takes a product
of two quaternions to describe a rotation (Appendix~\ref{app:A}). The
inverse relation
\begin{equation}\label{eq:phi2q}
\delta q_a = \frac{1}{2}\sum_{i=1}^3 e_{ia} \delta \phi_i
\end{equation}
automatically maintains the normalization of $q_a$. The angular
velocity and bending strains can be directly related to derivatives of
$q_a$,
\begin{equation}\label{eq:qderiv}
\omega_i = {\dot \phi}_i= 2e_{ia}{\dot q_a} ~~~ \Omega_i =
\phi_i^\prime = 2e_{iq}q_a^\prime.
\end{equation}

We are now in a position to write down ordinary differential equations
that approximate the dynamics of an elastic filament. A nice feature
of the midpoint discretization\cite{Dichmann1996} is that the strains
are naturally evaluated at integer multiples of the segment length, $n
\Delta s$, with $n = 0, 1, \ldots, N$. An additional differencing of
the internal forces and couples then gives accelerations back at the
nodal positions. Thus the algorithm is second-order accurate in
$\Delta s$, with only three nodes directly interacting with one
another, just as in the WLC model. The derivatives ${r_\alpha^{\prime
n}},{q_a^{\prime  n}}$ are approximated by centered differences at the
discrete locations $n \Delta s$, midway between the nodes,
\begin{eqnarray}
r_\alpha^{\prime n} = \frac{r_\alpha^{n+1}(t) - r_\alpha^n(t)}{\Delta
s} + {\cal O}(\Delta s)^2, \label{eq:rdiff} \\ q_a^{\prime  n} =
\frac{q_a^{n+1}(t) - q_a^n(t)}{\Delta s} + {\cal O}(\Delta
s)^2. \label{eq:qdiff}
\end{eqnarray}
In addition we need to estimate the quaternions at $n \Delta s$ in
order to calculate the rotation matrices,
Eqs.~\ref{eq:di}--\ref{eq:ei},
\begin{eqnarray}
{\bar q_a}^n = \frac{q_a^{n+1}(t) + q_a^n(t)}{\modulo{q_a^{n+1}(t) +
q_a^n(t)}} + {\cal O}(\Delta s)^2. \label{eq:qsum}
\end{eqnarray}
Thus the coordinates, $r_\alpha^n, q_a^n$, are evaluated at the nodal
positions, $(n + 1/2) \Delta s$, while the derivatives
$r_\alpha^{\prime,n}, q_a^{\prime,n}$, and mean, ${\bar q}_a^n$, are
evaluated at $n \Delta s$.

The elastic forces and couples at the interior positions $n \Delta s$,
$n = 1, 2, \ldots, N-1$, are then
\begin{eqnarray}
F_\alpha^{\Gamma,  n} &=& \sum_{i=1}^3 C_i^\Gamma {\bar d}_{i\alpha}^n
\left( {\bar d}_{i\beta}^n r_\beta^{\prime  n} - \Gamma_i^0 \right),
\label{eq:FGamma} \\ F_\alpha^{\Omega,  n} &=& \sum_{i=1}^3 C_i^\Omega
{\bar d}_{i\alpha}^n \left( 2{\bar e}_{ib}^n q_b^{\prime  n} -
\Omega_i^0 \right), \label{eq:FOmega}
\end{eqnarray}
where the notation ${\bar d}_{i\alpha}^n$ and ${\bar e}_{ia}^n$
indicates the basis vectors are calculated from the average
quaternions ${\bar q}_a^n$ (Eq.~\ref{eq:qsum}). The forces at the ends
of the rod, $n = 0$ and $n=N$, are determined by the boundary
conditions. For free ends,
\begin{equation}
F_\alpha^{\Gamma,  0} = F_\alpha^{\Gamma, N} = F_\alpha^{\Omega,  0} =
F_\alpha^{\Omega, N} = 0, \label{eq:free}
\end{equation}
while prescribed external forces and couples on the ends of the rod
can also be included. Dirichlet boundary conditions require virtual
nodes, $n=0$ and $n=N+1$, which are constructed to satisfy the
boundary conditions at the ends of the
filament.\cite{Dichmann1996} For example, if the position and
orientation of the rod at $s=0$ are specified by ${\bar r}_\alpha^0$
and ${\bar q}_a^0$, then the virtual coordinates are
\begin{eqnarray}
r_\alpha^0 &=& 2{\bar r}_\alpha^0 - r_\alpha^1, \\ q_a^0 &=&
\frac{2{\bar q}_a^0 - q_a^1}{\sqrt{(2{\bar q}_a^0- q_a^1)(2{\bar
q}_a^0 - q_a^1)}}.
\end{eqnarray}
The elastic forces and couples at $s = 0$ can then be determined in
the same way as for the interior nodes. However, it seems preferable
to implement Dirichlet conditions by placing the nodes at integer
locations along the filament, $n \Delta s$, and then calculating the
forces at the half-integer positions; this eliminates the need for
virtual nodes. In the case of mixed boundary conditions a combination
of these strategies may be necessary, depending on the specifics of
the problem; in this paper we just consider filaments with force and
couple free boundaries.

The nodal coordinates and momenta satisfy the ordinary differential
equations ($n=1,2, \ldots, N$)
\begin{eqnarray}
{\dot r}_\alpha^n &=& \frac{p_\alpha^n}{M^\Gamma}, \label{eq:DFr} \\
{\dot q}_a^n &=& \frac{1}{2}\sum_{i=1}^3 \frac{e_{ia}^n d_{i\alpha}^n
l_\alpha^n}{M_i^\Omega}, \label{eq:DFq}\\ {\dot p}_\alpha^n &=&
f_\alpha^n = \frac{F_\alpha^{\Gamma,  n}-F_\alpha^{\Gamma,
n-1}}{\Delta s}, \label{eq:DFp}\\ {\dot l}_\alpha^n  &=& t_\alpha^n =
\left( \frac{F_\alpha^{\Omega,  n}-F_\alpha^{\Omega,  n-1}}{\Delta s}
+ \sum_{i,j,k=1}^3 \epsilon_{ijk} d_{i\alpha}^n \frac{(\Gamma_j^n +
\Gamma_j^{n-1}) (F_k^{\Gamma,  n}+F_k^{\Gamma,  n-1})}{4}
\right). \label{eq:DFl}
\end{eqnarray} 
The rotation matrices $d_{i\alpha}^n$ and $e_{ia}^n$, without the
overbar (\cf Eqs.~\ref{eq:FGamma} and~\ref{eq:FOmega}), are evaluated
from the nodal quaternions $q_a^n$, whereas the strains $\Gamma_i^n$,
$\Omega_i^n$ and forces $F_i^{\Gamma,n}$, $F_i^{\Omega,n}$ are
evaluated at the points $n \Delta s$, midway between nodes $n$ and
$n-1$. The numerical approximation to the term $\VGamma \times
\VF^\Gamma$ requires nodal values of $\VGamma$ and $\VF^\Gamma$, which
are determined by averaging the body-fixed strains and forces, and
then rotating the vector product to the space-fixed frame
(Eq.~\ref{eq:DFl}).

\section{Hamiltonian formulation}\label{sec:H}

The standard procedure for solving the partial differential equations
for the linear and angular
momenta\cite{Simo1988,Simo1995,Romero2002,Bishop2004,Goyal2005} does
not, in general, lead to a symplectic algorithm, because the discrete
nodal forces are not derived from a potential energy function. Rather
than discretize the equations of motion for the continuum rod, we
instead discretize the line integral making up the Hamiltonian
function,\cite{Dichmann1996} to obtain a discrete Hamiltonian that is
a second order (in $\Delta s$) approximation to ${\cal H} = {\cal T} +
{\cal U}$. We then use time integration schemes that preserve the
symplectic structure of the discrete
Hamiltonian.\cite{Dichmann1996,Miller2002}

\subsection{Hamiltonian for an elastic filament}

The kinetic (Eq.~\ref{eq:T}) and potential (Eq.~\ref{eq:U}) energies
of an elastic filament can be written in terms of the coordinates and
their space and time derivatives,
\begin{eqnarray}
{\cal T} &=& \frac{1}{2} \int_0^L \left( M^\Gamma{\dot r}_\alpha {\dot
r}_\alpha + 4 \sum_{i=1}^3  M_i^\Omega e_{ia} e_{ib} {\dot q}_a {\dot
q}_b \right) ds, \label{eq:TP} \\ {\cal U}&=& \frac{1}{2} \int_0^L
\sum_{i=1}^3  \left[ C_i^\Gamma (d_{i\alpha} r_\alpha^\prime -
\Gamma_i^0)  (d_{i\beta} r_\beta^\prime - \Gamma_i^0) + C_i^\Omega (2
e_{ia} q_a^\prime - \Omega_i^0)(2 e_{ib} q_b^\prime - \Omega_i^0)
\right] ds. \label{eq:UQ}
\end{eqnarray}
The first step is to identify the momentum fields, $\VP = \partial
T/\partial {\dot \VQ}$, conjugate to our chosen coordinates, $\VQ(s,t)
= [r_\alpha(s,t), q_a(s,t)]$:
\begin{equation}
p_\alpha = M^\Gamma {\dot r}_\alpha, ~~~ l_a = 4 \sum_{i=1}^3
M_i^\Omega e_{ia} e_{ib} {\dot q}_b,
\end{equation} 
where $l_a = [l_0, l_x, l_y, l_z]$ is the angular momentum field
conjugate to $q_a$. It is related to the body-fixed angular momentum
field, $l_i = M_i^\Omega \omega_i = 2 M_i^\Omega e_{ib}{\dot q}_b$,
\begin{equation}
l_a = 2  \sum_{i=1}^3 l_i e_{ia}, ~~~ l_i = \frac{1}{2} e_{ia}
l_a. \label{eq:lali}
\end{equation}

Rewriting the kinetic energy in terms of the conjugate momenta,
\begin{equation}
{\cal T} = \frac{1}{2} \int_0^L \left( \frac{p_\alpha
p_\alpha}{M^\Gamma} + \frac{1}{4} \sum_{i=1}^3 e_{ia} e_{ib}\frac{l_a
l_b}{M_i^\Omega} \right) ds, \label{eq:TPQ}
\end{equation}
we can derive the equations of motion of the coordinates by functional
differentiation of ${\cal T(\VP,\VQ})$ with respect to $\VP$:
\begin{eqnarray}
{\dot r}_\alpha &=& \frac{\delta {\cal T}}{\delta p_\alpha} =
\frac{p_\alpha}{M^\Gamma}, \label{eq:Hr} \\ {\dot q}_a &=&
\frac{\delta {\cal T}}{\delta l_a} = \frac{1}{4}\sum_{i=1}^3
\frac{e_{ia} e_{ib} l_b}{M_i^\Omega} = \frac{1}{2} \sum_{i=1}^3
\frac{e_{ia} l_i}{M_i^\Omega}. \label{eq:Hq}
\end{eqnarray}

The equation of motion for the linear momentum field derives from the
potential energy due to shear and extension (Eq.~\ref{eq:UQ}),
\begin{equation}
{\dot p_\alpha} = - \frac{\delta U^\Gamma}{\delta r_\alpha} = -
\int_0^L F_\beta^\Gamma \frac{\delta r_\beta^\prime}{\delta r_\alpha}
ds. \label{eq:dUdr}
\end{equation}
The functional derivative requires an integration by parts to convert
variations in $\Vr^\prime$ to variations in $\Vr$,
\begin{equation}
{\dot p_\alpha} = \Ds F_\alpha^\Omega,  \label{eq:Hp}
\end{equation} 
as before (Eq.~\ref{eq:EOMp}). Here we have omitted contributions
derived from work done on the ends of the rod by external forces,
which we assume are included in an external interaction potential
$U^E$.

The angular momentum field has three contributions; from ${\cal T}$,
${\cal U}^\Gamma$, and ${\cal U}^\Omega$,
\begin{equation}
{\dot l_a} = \Ds F_a^\Omega + \sum_{i,j,k=1}^3 \epsilon_{ijk} e_{ia}
\left( \frac{-l_j l_k}{M_j^\Omega} + \Omega_j F_k^\Omega + 2 \Gamma_j
F_k^\Gamma \right) + 2 q_a \sum_{i=1}^3  \Gamma_i
F_i^\Gamma. \label{eq:Hla}
\end{equation}
The functional derivative of ${\cal U}^\Omega$ was evaluated following
Eq.~\ref{eq:dUdr}, but includes an additional term derived from the
rotation of the frame by variations in $q_a$. There are similar
contributions from rotations of the frame in the functional
derivatives of ${\cal T}$ and ${\cal U}^\Gamma$. Derivatives of the
basis vectors $d_{i\alpha}$ and $e_{ia}$ with respect to $q_a$ were
evaluated using Eqs.~\ref{eq:ddi}--\ref{eq:dei} from
Table~\ref{tb:q}. Although the equations of motion must be derived for
the canonical momenta $p_\alpha$ and $l_a$, the numerical
implementation can use any frame. We have found that it is most
convenient to use space-fixed linear momenta and body-fixed angular
momenta as the primary variables, since this seems to minimize the
number of rotations of $\Vl$. The quaternion momenta can be rewritten
as body-fixed momenta, ${\dot l_i} = ({\dot e_{ia}} l_a + e_{ia}{\dot
l_a})/2$,
\begin{equation}
{\dot l_i} + \sum_{j,k=1}^3 \epsilon_{ijk} \frac{l_j l_k}{M_j^\Omega}
= \Ds F_i^\Omega + \sum_{j,k=1}^3 \epsilon_{ijk} \left( \Omega_j
F_k^\Omega + \Gamma_j F_k^\Gamma \right),  \label{eq:Hl}
\end{equation}
again using Eq.~\ref{eq:dei} to evaluate variations in $e_{ia}$. This
expression is equivalent to Eq.~\ref{eq:EOMl} except that it is
written in the body-fixed frame instead of the space-fixed frame.

\subsection{Discretized Hamiltonian}\label{sec:DH}

In this section we will derive equations of motion for the nodal
coordinates and momenta by discretizing the line integrals in
Eqs.~\ref{eq:UQ} and~\ref{eq:TPQ}. The kinetic energy is approximated
by the midpoint rule,
\begin{equation}
{\cal T}^N = \frac{1}{2} \sum_{n=1}^N \left( \frac{p_\alpha^n
p_\alpha^n}{M^\Gamma} + \frac{1}{4} \sum_{i=1}^3 e_{ia}^n e_{ib}^n
\frac{l_a^n l_b^n}{M_i^\Omega} \right), \label{eq:TD}
\end{equation} 
where ${\cal T}^N$ is the discrete kinetic energy per unit length. The
discrete Hamiltonian of a set of infinitesimal segments, ${\cal H}^N$,
is an energy density, whereas a Hamiltonian describing finite-length
segments\cite{Butler2005,Qi2006} would have units of
energy. Equation~\ref{eq:TD} is a second order approximation to the
kinetic energy of the continuous filament, ${\cal T} = {\cal T}^N
\Delta s + {\cal O}(\Delta s)^3$. Discrete approximations to the
potential energy involve coordinate differences evaluated at the
midpoints between pairs of nodes. We therefore approximate the
potential energy by a trapezoidal rule, which is also second order in
$\Delta s$,
\begin{equation}
{\cal U}^N = \frac{1}{2} \sum_{n=0}^N \sum_{i=1}^3  w_n \left[
C_i^\Gamma ({\bar d}_{i\alpha}^n r_\alpha^{\prime n} - \Gamma_i^0)
({\bar d}_{i\beta}^n r_\beta^{\prime n}- \Gamma_i^0) + C_i^\Omega (2
{\bar e}_{ia}^n q_a^{\prime n} - \Omega_i^0)(2 {\bar e}_{ib}^n
q_b^{\prime n} - \Omega_i^0) \right]. \label{eq:UD}
\end{equation} 
The derivatives $r_\alpha^{\prime n}$ and $q_a^{\prime n}$ are defined
in Eqs.~\ref{eq:rdiff}--\ref{eq:qdiff} and the average quaternions
${\bar q}_{a}^n$, used to calculate ${\bar e}_{ia}^n$, are defined in
Eq.~\ref{eq:qsum}. The weights, $w_n$, for the trapezoidal integration
rule are $w_n = 1/2$ if $n = 0$ or $n = N$ and $w_n = 1$ otherwise.

The equations of motion for the nodal coordinates and momenta then
follow by differentiation:
\begin{eqnarray}
{\dot r}_\alpha^n &=& \frac{\partial  {\cal T}^N}{\partial p_\alpha^n}
= \frac{p_\alpha^n}{M^\Gamma}, \label{eq:DHr} \\ {\dot q}_a^n &=&
\frac{\partial  {\cal T}^N}{\partial l_a^n} = \frac{1}{2}\sum_{i=1}^3
\frac{e_{ia}^n l_i^n}{M_i^\Omega}, \label{eq:DHq}\\ {\dot p}_\alpha^n
&=& -\frac{\partial  {\cal U}^N}{\partial r_\alpha^n} = f_\alpha^n,
\label{eq:DHp}\\ {\dot l}_a^n  &=& -\frac{\partial  {\cal
H}^N}{\partial q_a^n} = - \sum_{i,j,k=1}^3 \epsilon_{ijk}
\frac{e_{ia}^n  l_j^n l_k^n}{M_j^\Omega} + t_a^n, \label{eq:DHla}
\end{eqnarray}
where the nodal forces and torques are
\begin{eqnarray}
f_\alpha^n &=& \frac{w_n F_\alpha^{\Gamma,  n} - w_{n-1}
F_\alpha^{\Gamma, n-1}}{\Delta s}, \label{eq:DHfC} \\ t_a^n &=&
\frac{w_n F_a^{\Omega,  n}-w_{n-1} F_a^{\Omega,  n-1}}{\Delta s}
\nonumber \\ &+&  \sum_{i,j,k=1}^3 \epsilon_{ijk} \left( w_n{\bar
e}_{ia}^n \frac{\Omega_j^n F_k^{\Omega, n}}{2{\bar q}^n} +
w_{n-1}{\bar e}_{ia}^{n-1} \frac{\Omega_j^{n-1} F_k^{\Omega,
n-1}}{2{\bar q}^{n-1}} \right)  \nonumber \\ &+&  \sum_{i,j,k=1}^3
\epsilon_{ijk} \left( w_n{\bar e}_{ia}^n \frac{\Gamma_j^n F_k^{\Gamma,
n}}{{\bar q}^n} + w_{n-1}{\bar e}_{ia}^{n-1} \frac{\Gamma_j^{n-1}
F_k^{\Gamma,  n-1}}{{\bar q}^{n-1}} \right), \label{eq:DHtC}
\end{eqnarray}
and ${\bar q}^n$ is the length of the unnormalized quaternion ${\bar
q}^n = \modulo{q_a^n + q_a^{n-1}}/2$. It is essential that the
differentiation is done exactly, otherwise the Hamiltonian structure
of the equations of motion is
lost. Equations~\ref{eq:DHr}--\ref{eq:DHfC} are straightforward, but
Eq.~\ref{eq:DHtC} requires some explanation. The factor of two between
the $\VGamma \times \VF^\Gamma$ and $\VOmega \times \VF^\Omega$
contributions (\cf Eq.~\ref{eq:Hla}) arises because the rate of
rotation of the quaternion basis is one-half that of the body-fixed
frame. Terms involving dot products of ${\bar q_a}^n$ with ${\bar
e_{ia}}^n$ vanish by orthogonality, even for the midpoint
quaternions. Less obviously, the orthogonality of $q_a$ and
$q_a^\prime$ is preserved by the discretization, so that
\begin{equation}
{\bar q_a}^n q_a^{\prime n} = \left( \frac{q_a^{n+1}+q_a^n}{2} \right)
\left( \frac{q_a^{n+1}-q_a^n}{\Delta s} \right) = 0.
\end{equation} 
Although the discrete Hamiltonian, ${\cal H}^N = {\cal T}^N + {\cal
U}^N$, is only a second-order approximation to ${\cal H}$, the
equations of motion  for the nodes (Eqs.~\ref{eq:DHr}--\ref{eq:DHtC})
exactly preserve a Hamiltonian structure for any $\Delta
s$. Equations~\ref{eq:DFq}--\ref{eq:DFl} do not have this property,
although they are the same to second order in $\Delta s$.

For our numerical implementation, it is more convenient to calculate
the angular momenta in the body-fixed frame rather than the quaternion
basis. Making the same transformation as from Eq.~\ref{eq:Hla} to
Eq.~\ref{eq:Hl},
\begin{equation}
{\dot l_i^n} + \sum_{j,k=1}^3 \epsilon_{ijk} \frac{l_j^n
l_k^n}{M_j^\Omega} = \frac{1}{2} e_{ia}^n t_a^n ,  \label{eq:DHl}
\end{equation}
where the conservative torque in the quaternion basis is given by
Eq.~\ref{eq:DHtC}. No further simplification is possible in this case,
because the quaternion basis vector $e_{ia}^n$ is not the same as
those in the expression for $t_a^n$. The slight variations in the
quaternions make the difference between the Hamiltonian formulation
for the torque (Eq.~\ref{eq:DHtC}) and the torque (Eq.~\ref{eq:DFl})
derived from the finite-difference discretization described in
Sec.~\ref{sec:DEOM}.

\subsection{Operator splitting}\label{sec:OS}

Implicit integration methods are typically used to integrate the
equations of motion of elastic
rods,\cite{Dichmann1996,Romero2002,Goyal2005} even when the model has
no explicit constraints.\cite{Romero2002} The most common choice is
the implicit midpoint method, which updates the vector $\VY =
[\VP,\VQ]$ to second order in the time step $\Delta t$,
\begin{equation}\label{eq:MP}
\VY(t+\Delta t) = \VY(t) + \frac{\Delta t}{2}\left({\dot \VY}[\VY(t)]
+ {\dot \VY}[\VY(t+\Delta t)] \right).
\end{equation}
Implicit methods have the advantage of stability for large time steps
and the implicit midpoint method is in addition
symplectic.\cite{McLachlan1992} However a number of force evaluations
are needed at each time step to solve the non-linear
equations~\eqref{eq:MP} to machine precision, which is necessary to
maintain the symplectic structure. Moreover, the normalization
constraint on the quaternion is not conserved,
\begin{equation}
\modulo {q_a^{k+1}} = 1 + \frac{h^2}{16}\sum_{i=1}^3 \left(
\omega_i^{k+1} \right)^2 - \left( \omega_i^k \right)^2,
\end{equation}
and must be rescaled at each time step.

Operator splitting techniques are increasingly being used to solve
both deterministic\cite{Dullweber1997,Miller2002,Zon2008} and
stochastic differential
equations.\cite{Serrano2006,Fabritiis2006} Typically the splitting is
devised so that the individual propagators can be determined
exactly. If the underlying dynamics is strictly
Hamiltonian,\cite{Dullweber1997,Miller2002,Zon2008} then symplectic
integrators can be constructed by such techniques. The Liouville
operator, ${\cal L}= {\cal L}^T + {\cal L}^U$, is decomposed into
kinetic (${\cal L}^T$) and potential (${\cal L}^U$) terms,
\begin{eqnarray}
{\cal L}^T &=& \sum_{n=1}^N \left( {\dot r_\alpha^n}
\frac{\partial}{\partial r_\alpha^n} + {\dot q_a^n}
\frac{\partial}{\partial q_a^n} \right), \label{eq:DLT} \\ {\cal L}^U
&=& \sum_{n=1}^N \left( f_\alpha^n \frac{\partial}{\partial
p_\alpha^n} + t_\alpha^n \frac{\partial}{\partial l_\alpha^n} \right):
\label{eq:DLF}
\end{eqnarray}
here we use a second-order Trotter
decomposition,\cite{Dullweber1997,Miller2002}
\begin{equation}
\exp \left[ {{\cal L} \Delta t} \right] = \exp \left[ {{\cal L}^T
\Delta t/2} \right] \exp \left[{\cal L}^U \Delta t \right] \exp \left[
{\cal L}^T \Delta t/2 \right] + {\cal O}(\Delta t)^3,
\label{eq:Trotter}
\end{equation} 
although higher-order algorithms are
available.\cite{Omelyan2002,Omelyan2003}

The integration of the position and momentum equations is a
straightforward and exact streaming,
\begin{eqnarray}
r_\alpha(\Delta t) &=&  \exp \left[ ({\cal L}^T \Delta t \right]
r_\alpha = r_\alpha + \frac{p_\alpha}{M^\Gamma} \Delta t,
\label{eq:FSr} \\ p_\alpha^n(\Delta t) &=&  \exp \left[ {\cal L}^U
\Delta t \right] p_\alpha^n = p_\alpha^n + f_\alpha^n \Delta t,
\label{eq:FSp} \\ l_i^n(\Delta t) &=&  \exp \left[ {\cal L}^U \Delta t
\right] l_i^n = l_i + t_i^n \Delta t. \label{eq:FSl}
\end{eqnarray}
An exact solution of the quaternion update is more complicated, but
can be carried out using elliptic
integrals.\cite{Zon2008} Nevertheless, here we adopt a simpler
formulation which uses a sequence of rotations about the body-fixed
axes,
\begin{equation}
{\cal L}^T = \sum_{n=1}^N \left( {\dot r_\alpha^n}
\frac{\partial}{\partial r_\alpha^n} + \sum_{i=1}^3 {\cal L}_i^n
\right), ~~~ {\cal L}_i^n = \frac{l_i^n}{2 M_i^\Omega} e_{ia}^n
\frac{\partial}{\partial q_a^n}. \label{eq:DLLi}
\end{equation}
A rotation $\Delta \phi_i^n = l_i^n \Delta t /M_i^\Omega$ about one of
the body-fixed axes changes both the quaternions and the other
body-fixed momenta:
\begin{eqnarray}
\exp \left({\cal L}_i^n \Delta t \right) q_a^n &=& \cos (\Delta
\phi_i^n /2) q_a^n + \sin  (\Delta \phi_i^n /2) e_{ia}^n,
\label{eq:DLq} \\ \exp \left({\cal L}_i^n \Delta t \right) l_j^n &=&
\cos (\Delta \phi_i^n)l_j^n + \sum_{k=1}^3 \epsilon_{ijk} \sin (\Delta
\phi_i^n) l_k^n. \label{eq:DLl}
\end{eqnarray}
The individual rotations can be combined using any  suitable
second-order decomposition for $\sum_{i=1}^3 {\cal L}_i^n$, for example
\begin{equation} 
\left( \exp \left[ {{\cal L}_1^n \Delta t/2J} \right] \exp \left[
{{\cal L}_2^n \Delta t/2J} \right] \exp \left[ {{\cal L}_3^n \Delta
t/J} \right] \exp \left[ {{\cal L}_2^n \Delta t/2J} \right] \exp
\left[ {{\cal L}_1^n \Delta t/2J} \right] \right)^J.
\end{equation} 
The update of the quaternions is not exact, but it is symplectic and
exactly preserves the norm of the quaternion. If the time step is
broken up into $J$ subintervals, a more accurate integration can be
achieved without substantial overhead, since no force evaluation is
needed.\cite{Miller2002}

\section{Numerical examples}\label{sec:example}

Our analysis has been supplemented by numerical simulations using the
algorithms described in the text. We have compared explicit
fourth-order Runga-Kutta (RK) integration, implicit second-order
midpoint (MP) integration, and second-order Operator Splitting (OS)
(Sec.~\ref{sec:OS}). We have tried each method with forces and torques
derived from discretizing the partial differential equations (DF),
Eqs.~\eqref{eq:DFp}-\eqref{eq:DFl}, and with forces and torques
derived from discretizing the Hamiltonian (DH),
Eqs.~\eqref{eq:DHfC}-\eqref{eq:DHtC}. We investigated the stability
and conservation of energy from two initial conditions: a straight
filament bent into a circle and a straight filament bent into a helix.

\subsection{A filament bent into a circle}

\begin{figure}
\center \includegraphics[width=\fullwidth]{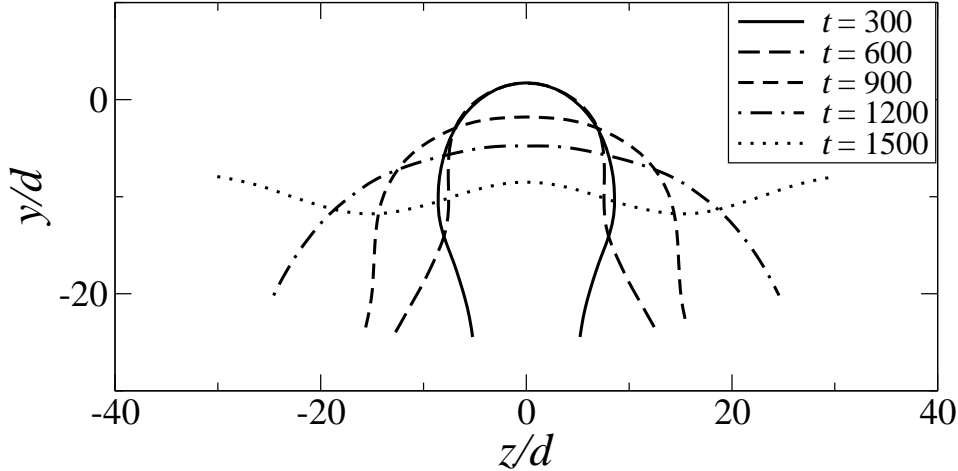}
\caption{Filament shapes at different times: $300 t_0$ (solid), $600
t_0$ (long dashes), $900 t_0$ (dashes), $1200 t_0$ (dot dash), and $1500
t_0$ (dotted). The time scale $t_0 = d/c_l$ is the time for a
longitudinal wave to cross the diameter of the filament}
\label{fig:shapes}
\end{figure} 

A straight filament of length $20 \pi d$ was bent into a circle of
radius $10 d$ and released. The dynamics were followed for two
different spatial discretizations, dividing the filament into $63$ or
$127$ equal segments; the corresponding segment lengths were
approximately $d$ and $0.5d$. The largest time step for the explicit
integrators is Courant limited by the time, $t_C$, for a longitudinal
wave to cross the shorter of the diameter, $d$, and the segment
length, $\Delta s$; we typically use a time step $\Delta t = 0.2
t_C$. As the rod evolves from its initial configuration, flexural
waves propagate along the filament, leading to a surprising variety of
configurations; a sampling of the filament shapes is illustrated in
Fig.~\ref{fig:shapes}. Initially the ends move slowly, and the
filament assumes a teardrop shape  ($t = 300 t_0$), followed by a
hairpin ($t = 600 t_0$) as the ends of the filament accelerate. The
time unit $t_0 = d/c_l$, where $c_l$ is the longitudinal wave
speed. The inverted U shape ($t = 900 t_0$) straightens out  ($t =
1200 t_0$), and then develops a "double-minimum" shape ($t = 1500
t_0$). The center of the filament moves down to complete the inversion
and the filament approximately retraces the sequence of shapes in
reverse order, to arrive at the inverted configuration at roughly half
the period of the main oscillation. However, the motion is not exactly
periodic because of the strong coupling between the flexural
modes. The interaction of flexural waves can lead to large local
stresses, exceeding that of the initial configuration; for example at
the top of the teardrop ($t = 300 t_0$) and at the bends in the
hairpin ($t = 600 t_0$). It has been shown that flexural modes can
cause unexpected fractures by this mechanism.\cite{Audoly2005}

 \begin{figure}
\center \includegraphics[width=\fullwidth]{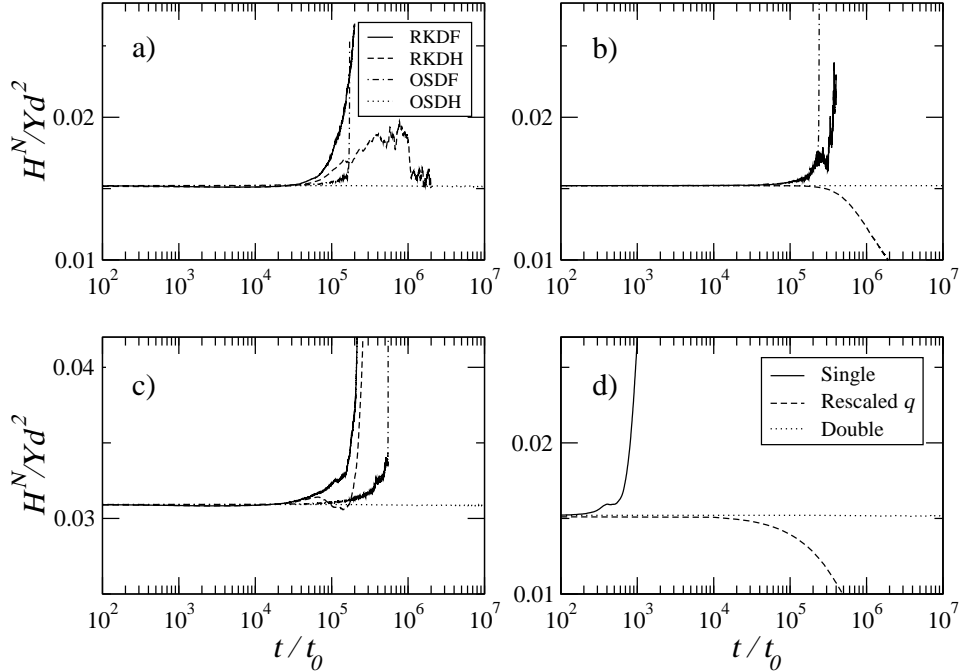}
\caption{Conservation of energy for symplectic (OSDH) and
non-symplectic (RKDF, RKDH, OSDF) algorithms. The initially circular
configuration of the filament unwinds as illustrated by the snapshots
in Fig.~\ref{fig:shapes}: a) 63 segments, $\Delta t = 0.2 t_0 = 0.2
t_C$; b) 63 segments, $\Delta t = 0.02 t_0 = 0.02t_C$; c) 127
segments, $\Delta t = 0.1 t_0 = 0.2 t_C$; d) OSDH algorithm with
varying precision, 63 segments, $\Delta t = 0.2 t_0 = 0.2 t_C$.}
\label{fig:energy}
\end{figure}

A complete cycle of the filament motion, back to a rough approximation
of its initial configuration, takes about $6000 t_0$ for a filament of
length $L \sim 60 d$, and is quadratic in the length of the
filament. The scaling is due to the dispersion relation of flexural
waves, $\omega \propto k^2$, which is quadratic rather than linear in
the wavevector ($k$); the period of the longest flexural wave, $8
\pi/(c_l k^2 d)$ is roughly $10^4 t_0$. A plot of energy \vs time,
Fig.~\ref{fig:energy}a, shows that all the algorithms integrate stably
for about 10 oscillations, but only the symplectic methods, MPDH and
OSDH, are stable at long times; on the scale of
Fig.~\ref{fig:energy}, results for MPDH and OSDH superpose, so only
the results for OSDH are shown. We have run the MPDH and OSDH
algorithms to a time of $10^8 t_0$ or $16000$ periods, with no
indication of instability. By contrast, changing the forces to the
non-Hamiltonian form (OSDF) or switching to the RK4 integrator (RKDH)
causes instabilities at times of the order of $10^5 t_0$. Reducing the
time step, Fig.~\ref{fig:energy}b, improves the stability of the
Runga-Kutta integration of the Hamiltonian forces (RKDH), increasing
the range of stability by about an order of magnitude. This is because
RKDH becomes symplectic in the limit $\Delta t \rightarrow 0$. On the
other hand if the forces are not Hamiltonian, reducing the time step
does not improve the stability; both RKDF and OSDF algorithms become
unstable after a time of about $10^5 t_0$, regardless of time
step. The discretized forces approach a Hamiltonian form in the limit
$\Delta s \rightarrow 0$ and reducing the segment length improves the
stability of the OSDF algorithm, extending the range of stability by
about a factor of 4 for a twofold reduction in the segment length,
Fig.~\ref{fig:energy}c. However, this is a double limiting process
requiring a progressively smaller time step as well as a reduced
segment length, making it computationally expensive. The RKDF
algorithm is not helped by a reduction in segment length; it
needs a further reduction in time step as well to see any improvement.

The non-linearity of the dynamics causes the filament to eventually
reach a state of thermal equilibrium, fluctuating around the straight
configuration. For the $63$ segment rod the equilibration time is
about $10^7 t_0$ independent of time step. For a constant filament
length, we observe that the equilibration time is roughly quadratic in
the number of segments. Thus the  behavior of this system in the
continuum limit is an interesting question for future work, but beyond
the scope of the present paper.

The stability of the symplectic integrator is affected by accumulated
round-off error. The results in Fig.~\ref{fig:energy}d show that the
symplectic integration scheme (OSDH) is quite unstable in single
precision arithmetic. The most rapid instability, at $t < 10^3 t_0$,
was traced to accumulated errors in the quaternion normalization. The
operator splitting algorithm maintains the quaternion normalization to
machine precision and with 64-bit arithmetic the normalization error
is stable at less than one part in $10^{14}$. But in single precision,
the error increases rapidly, which causes an incompatibility with the
assumption that the nodal quaternions are normalized. More puzzling is
that rescaling the quaternions does not solve the problem, but merely
delays the onset of the instability. However, if the initial
accumulation of round-off error is random, we would expect the double
precision version to run stably for about $10^{16}$ times longer, or
$10^{18} t_0$ which is well beyond the event horizon of the simulation.

\begin{figure}
\center \includegraphics[width=\fullwidth]{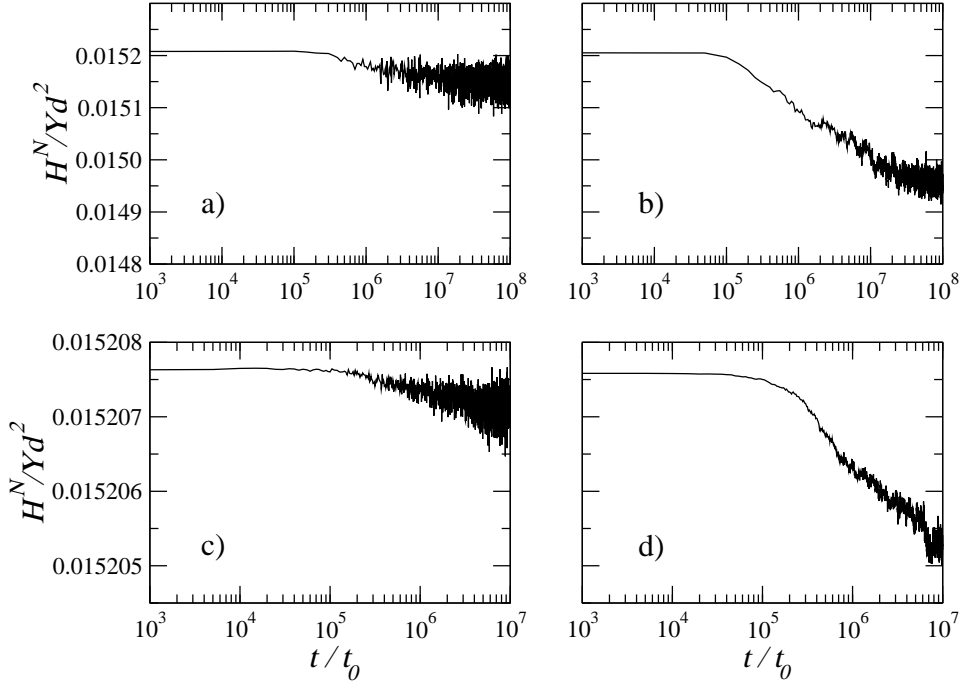}
\caption{Conservation of energy for symplectic algorithms OSDH and
MPDH; 63 segments were used in each case. a) OSDH, $\Delta t = 0.2
t_0$; b) MPDH $\Delta t = 0.2 t_0$; c) OSDH, $\Delta t = 0.02 t_0 $;
d) MPDH, $\Delta t = 0.02 t_0$.}
\label{fig:OSvsMP}
\end{figure}

The short-time fluctuations in energy of the OSDH algorithm cannot be
seen on the scale of Fig.~\ref{fig:energy}, but they are quadratic in
the time step, with a relative magnitude of approximately $0.1 (\Delta
t/t_0)^2$. These short-time fluctuations in energy are
about 20 times larger with OSDH than with MPDH. However there is also a
drift in the energy with time, again quadratic in $\Delta t$, but larger, as shown
in Fig.~\ref{fig:OSvsMP}. Over long time intervals, OSDH preserves
energy conservation with about an order of magnitude better accuracy than
MPDH at the same $\Delta t$ (Fig.~\ref{fig:OSvsMP}). MPDH requires 5-10
times as many force evaluations as OSDH per time step, so that the explicit operator
splitting algorithm is clearly preferable for long-time dynamics.

Dichmann and Maddocks studied the dynamics of a Kirchoff rod from the
same initial configuration,\cite{Dichmann1996} but with the filament
pinned at one end. The nodal forces and torques were also 
Hamiltonian, but the implicit midpoint integrator was used instead of
operator splitting. Their results showed a small drift in the total
energy of around $0.2\%$ after approximately 30 oscillations of the
filament, or $200,000 t_0$ in our units. Our results for the MPDH
algorithm behave in a qualitatively similar fashion; with a time step
$\Delta t = 0.2t_0$ we observe an accumulated energy drift of $0.3\%$
at $t = 200,000 t_0$. The error with OSDH is about an order of
magnitude smaller. The GE model requires a smaller time step to
explicitly integrate the shear and extensional degrees of freedom,
but surprisingly, it is only a factor of 8 smaller than the time step
used for the constrained rod.\cite{Dichmann1996} This suggests that the explicit OSDH
algorithm can integrate the full GE rod model with about the same
computational cost as an implicit integration of the Kirchoff
model. If excluded volume interactions are included, it is likely that
these very stiff forces will set the overall time step, as is typical
in molecular dynamics simulations. In such cases the computational
advantages of a fully explicit simulation will be considerable.

\subsection{A filament bent into a helix}

\begin{figure}
\center{\includegraphics[width=\halfwidth]{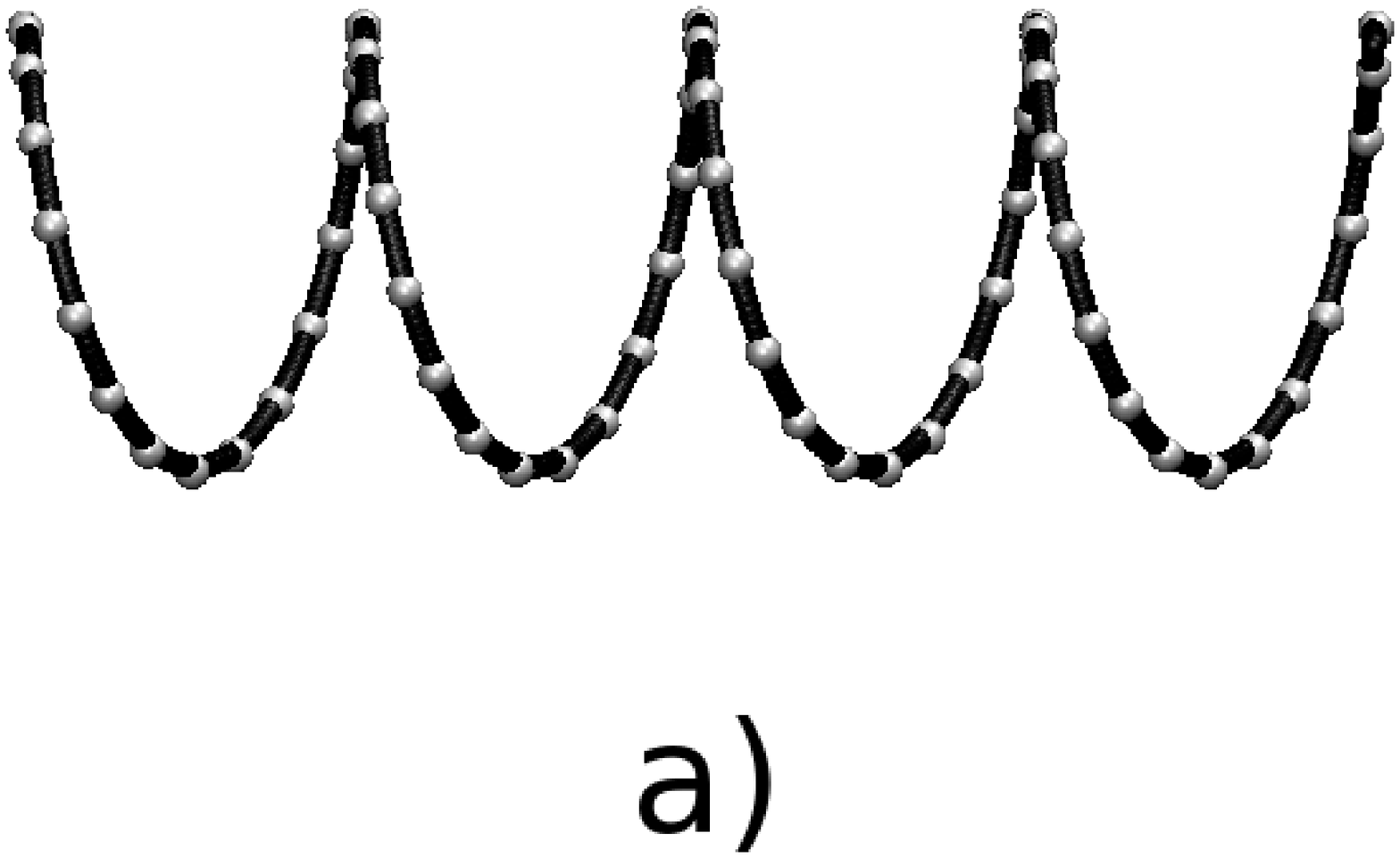}
	\includegraphics[width=\halfwidth]{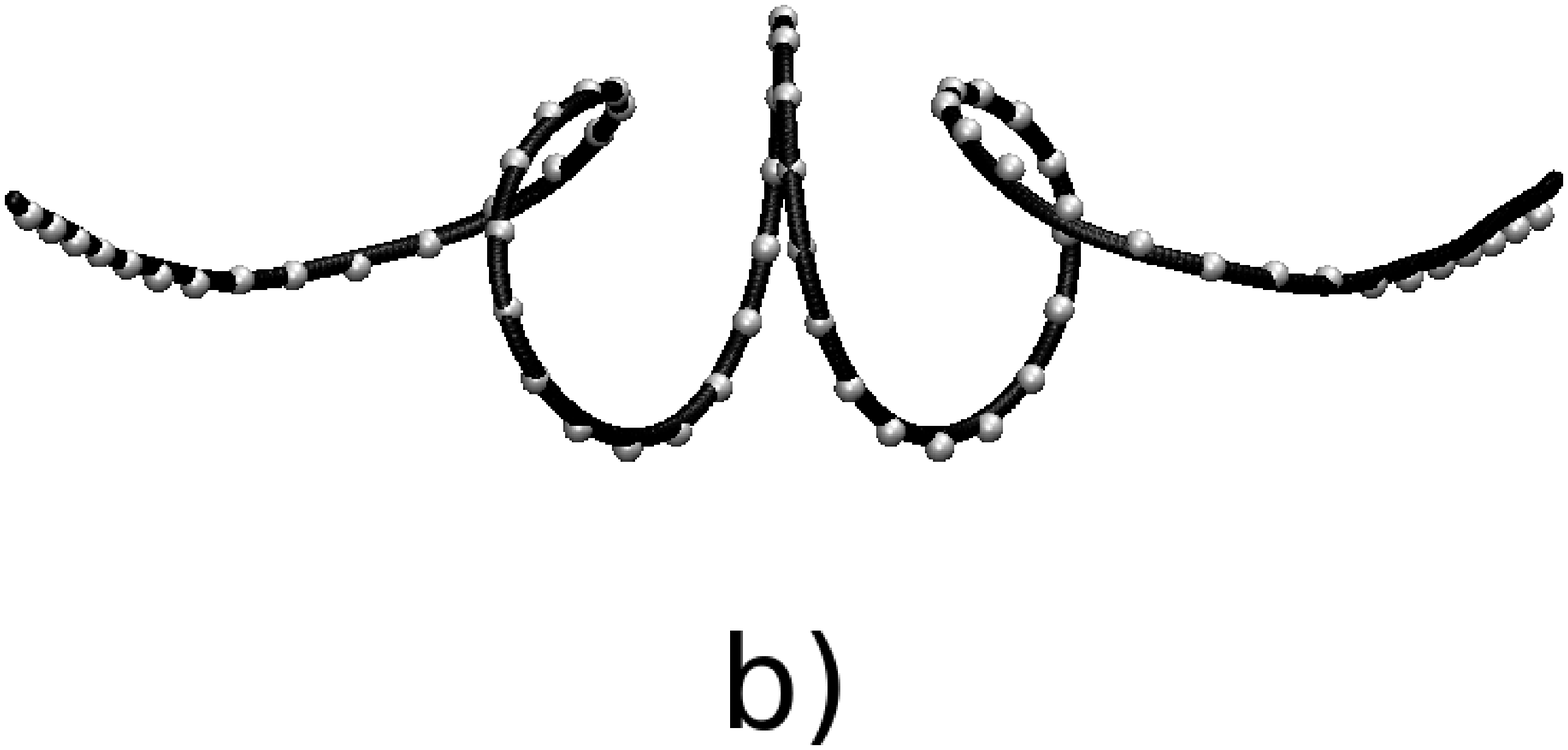}}
	\center{\includegraphics[width=\halfwidth]{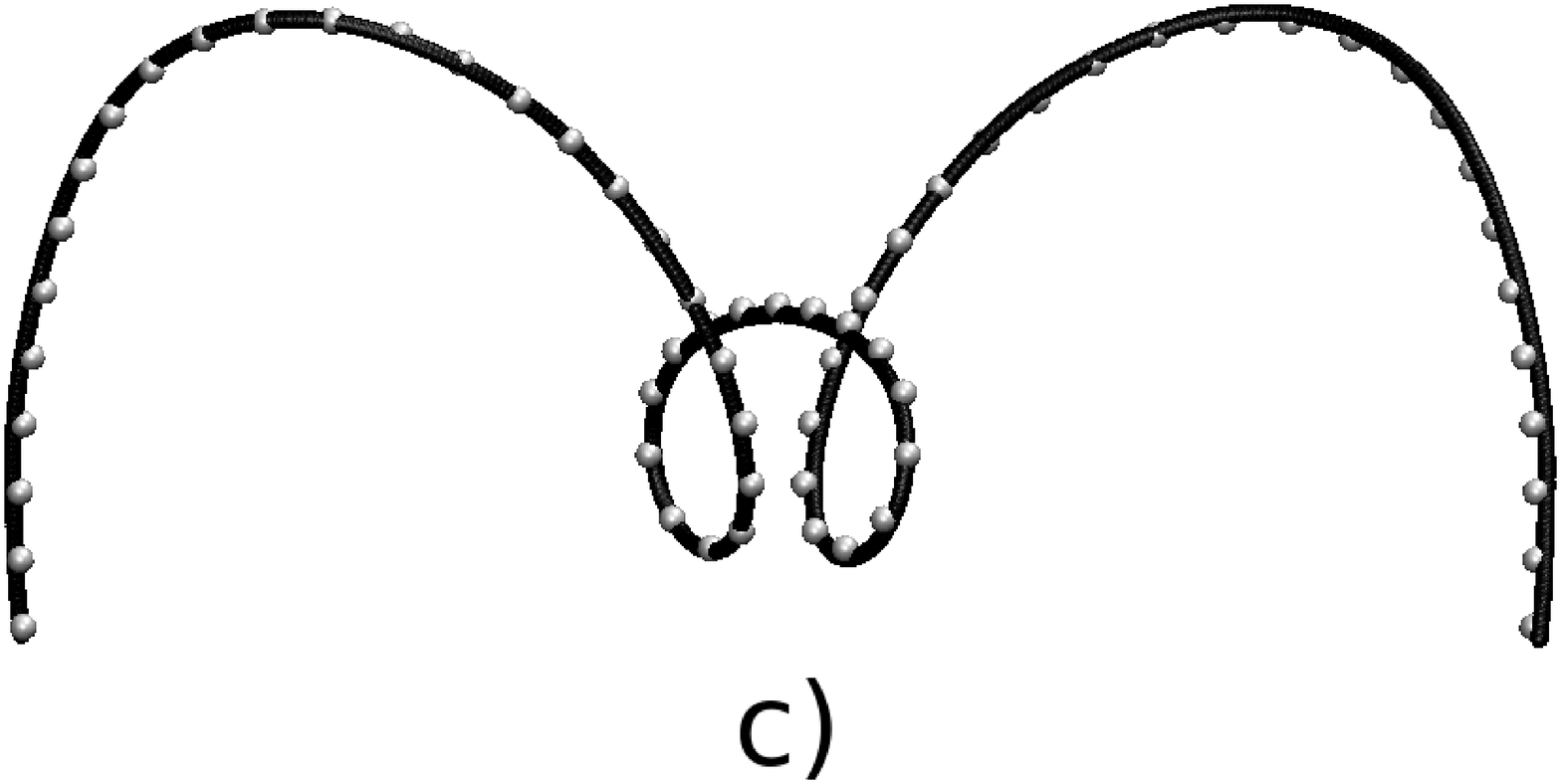}
	\includegraphics[width=\halfwidth]{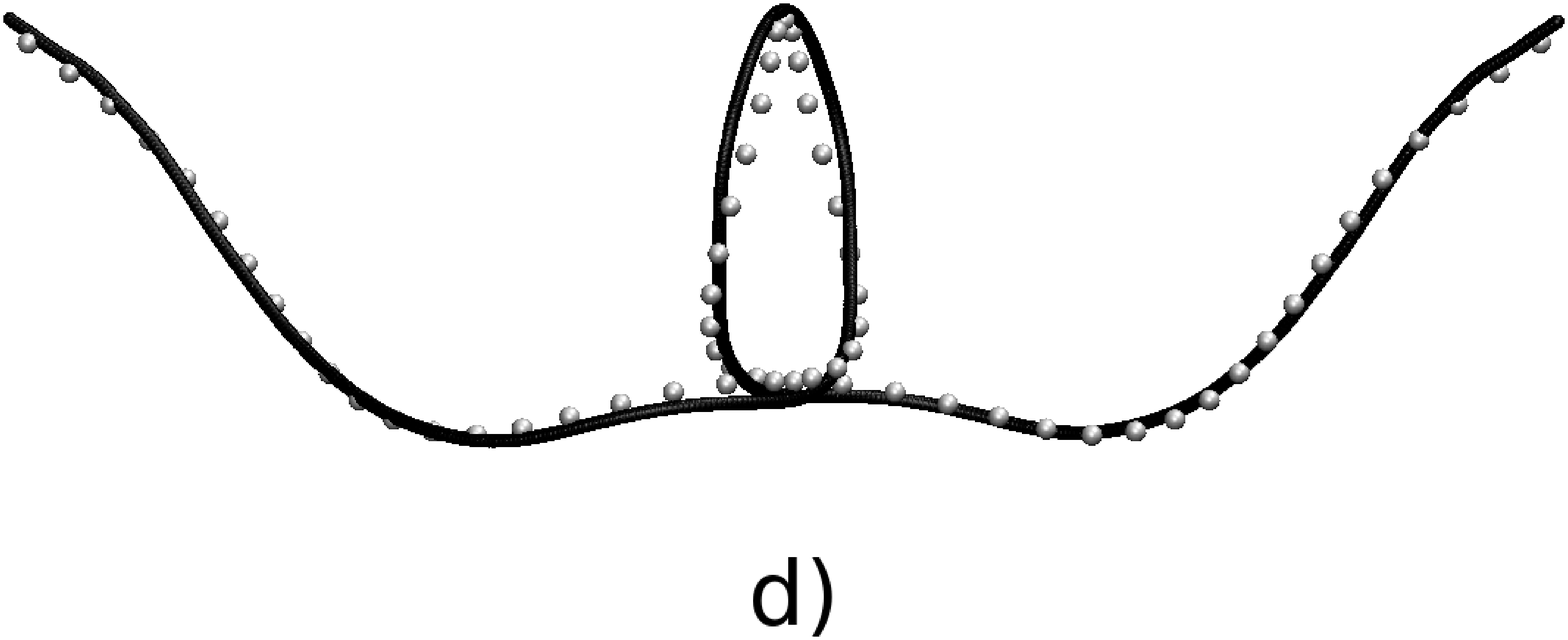}}
	\center{\includegraphics[width=\halfwidth]{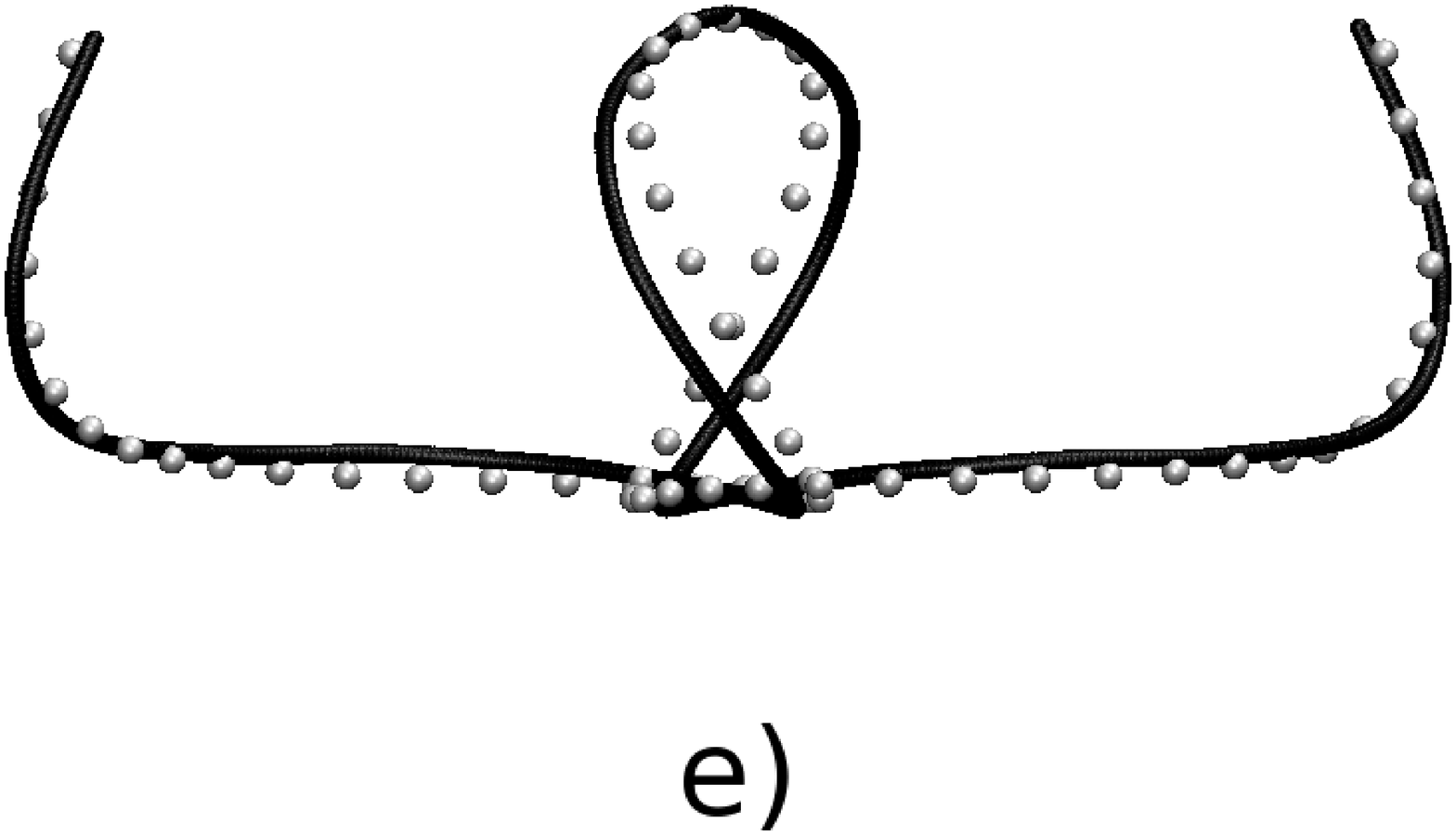}
	\includegraphics[width=\halfwidth]{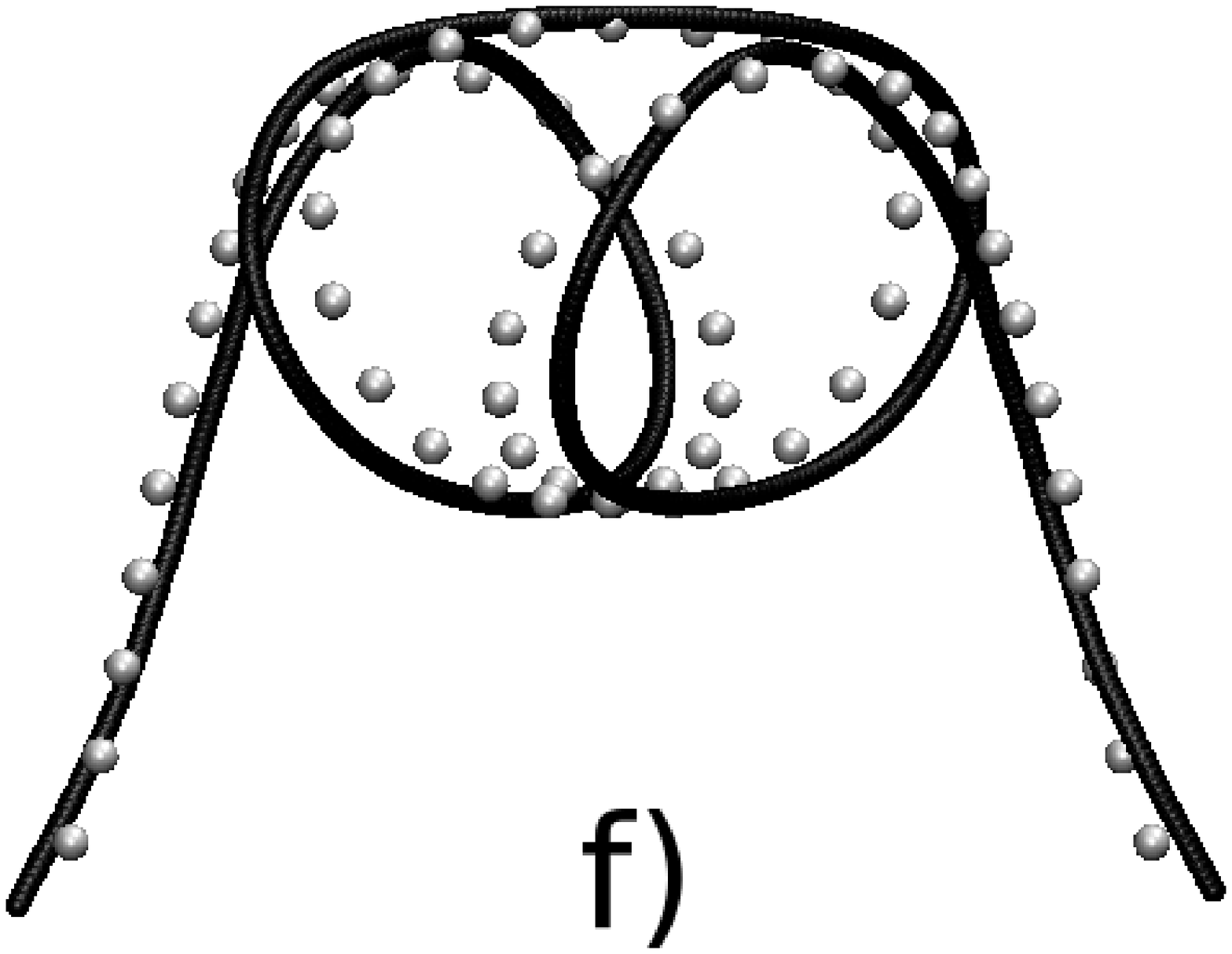}}
\caption{Filament shapes at different times: a) $t = 0$; b) $t =
100t_0$; c) $t = 200t_0$; d) $t = 300t_0$; e) $t = 400t_0$; f) $t =
500t_0$. The simulations with 630 segments are shown as thick solid
lines, while simulations with 63 segments are shown by the spheres.}
\label{fig:helix}
\end{figure}

We have also examined a more complicated initial condition, a straight
rod of length $20 \pi d$ wound into a tight helix with exactly four
complete turns. The curvature, $\VOmega = [0.4d^{-1},0,0.1d^{-1}]$, is
high and generates motion in all three spatial dimensions, which poses
a difficult challenge for the numerical method. We used two different
discretizations, $63$ segments of length $\Delta s \approx d$ and
$630$ segments of length $\Delta s \approx 0.1d$; snapshots of the
initial evolution of the filament shapes are shown in
Fig.~\ref{fig:helix}. There is a high degree of dynamical coherence
between the results at the two different resolutions, although the
strong nonlinearity of the problem means that they start to diverge at
times of the order of $500 t_0$. We did not include any excluded
volume interactions in these simulations, and the filaments can
therefore cross; this does not affect the accuracy of the numerical
algorithm.

\begin{figure}
\center \includegraphics[width=\fullwidth]{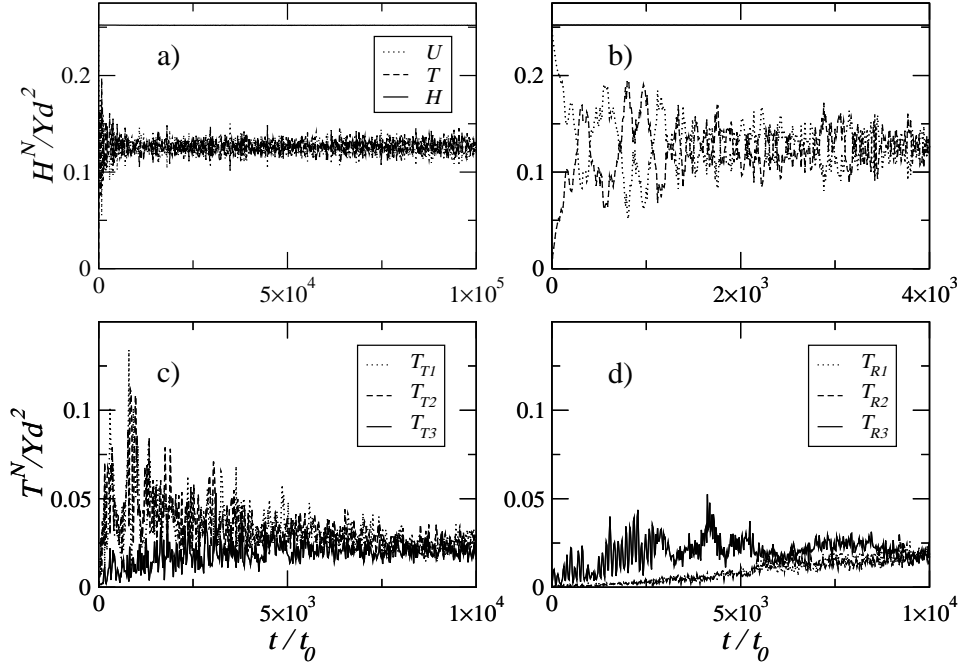}
\caption{Conservation of energy and thermal equilibrium with the
symplectic integrator OSDH. The initially helical configuration of the
filament unwinds as illustrated by the snapshots in
Fig.~\ref{fig:helix}. The kinetic, potential and total  energy of the
$63$ segment model ($\Delta t = 0.1 t_0 = 0.1 t_C$) are shown for: a)
$10^5 t_0$ and b) $4000t_0$. The body-fixed kinetic energy of the individual
degrees of freedom is also shown: c) Shear and extension and d)
Bending and torsion.}
\label{fig:hxenergy}
\end{figure}

As in the planar bend case, the symplectic algorithm (OSDH) conserves
energy, Fig.~\ref{fig:hxenergy}a, for as long a time as we have
tested, up to $10^6 t_0$. The non-linear coupling is much stronger
than in the previous example, because of the higher curvature and the
three-dimensional deformation; here the filament rapidly comes to
thermal equilibrium. The loss of coherent oscillations can be seen
more clearly in the expanded time scale of
Fig.~\ref{fig:hxenergy}b. Over the same time scale, $10^4 t_0$, we see
that equipartition of energy is established between the various
degrees of freedom, Figs.~\ref{fig:hxenergy}c and
~\ref{fig:hxenergy}d; similar results holds for the various components
of the potential energy as well. Unlike the planar bend case, here the
more finely resolved filament ($630$ segments) comes to thermal
equilibrium on more or less the same time scale, $\sim 40,000t_0$,
rather than $10^6 t_0$ as would be expected for a quadratic scaling of
the equilibration time with $N$. This suggests fundamental differences
in the dynamics of the two-dimensional bending from the full
three-dimensional problem.

\section{Conclusions}\label{sec:conc}

In this paper we have presented a new algorithm for simulating the
dynamics of elastic filaments. The test problems show the method to be
extremely stable, with exact conservation of momentum and angular
momentum (to machine precision), and global energy conservation to
order $\Delta t^2$. The algorithm is fully explicit and requires no
constraints of any kind, neither on the forces nor on the
quaternions. It is thus simpler in some ways than typical WLC
implementations which include extensional forces as a constraint. In
contrast to the WLC, the GE model correctly incorporates large bending
deformations and twisting; it includes the Kirchoff rod as a limiting
case.

Symplectic integration of the GE model can use a large time
step, within a factor of 10 of a constrained filament\cite{Dichmann1996}
that excludes shear and extensional modes. Explicit operator
splitting has better long-term energy conservation than the implicit
midpoint method and requires an order of magnitude fewer force
evaluations per time step. In cases where the time step is limited by
the stiffness of excluded volume interactions, the GE model may be
more computationally efficient than the Kirchoff model, due to the
absence of constraints.

In this work we only discussed Hamiltonian systems, but operator
splitting is a powerful method for integrating stochastic systems as
well.\cite{Serrano2006,Fabritiis2006} We have considered the case
when the rod is subjected to dissipative and random forces, in
addition to the elastic forces. Using operator splitting we can
integrate the momentum equation exactly, using the Ornstein-Uhlenbeck
solution, and therefore preserve quadratic norms to order $\Delta
t^2$, as opposed to the $\Delta t$ accuracy of Brownian dynamics. This
work will be reported in a future paper.

\acknowledgments{This work was supported by the National Science
Foundation (CTS-0505929) and the Alexander von Humboldt Foundation.}

\appendix
\section{Properties of quaternions}\label{app:A}

A quaternion ${\cal Z} = q_0 + q_x i + q_y j + q_z k$ is a complex
number with multiplicative identities
\begin{equation}\label{eq:quatdef}
i^2 = j^2 = k^2 = ijk = -1.
\end{equation}
We use the notation ${\cal Z}$ to indicate the quaternion and $q_a$ to
denote a vector containing the scalar, $q_0$, and vector, $\Vq =
[q_x,q_y,q_z]$, components of ${\cal Z}$.  The quaternion algebra,
Eq.~\eqref{eq:quatdef}, leads to rules for multiplication that are
analogous to the cross-product of unit vectors:
\begin{eqnarray}
ij = -ji = k \nonumber \\ jk = -kj = i \label{eq:quatmult} \\ ki = -ik
= j \nonumber
\end{eqnarray}
If we then identify $i$, $j$, $k$, with Cartesian unit vectors $\Vi$,
$\Vj$, $\Vk$, the multiplication of two quaternions, ${\cal Z} = q_0 +
q_x \Vi + q_y \Vj + q_z \Vk$ and ${\cal Z}^\prime = q_0^\prime +
q_x^\prime \Vi + q_y^\prime \Vj + q_z^\prime \Vk$, can be written,
using the quaternion multiplication rules defined in
Eqs. \ref{eq:quatdef} and \ref{eq:quatmult}, as
\begin{equation}\label{eq:quatprod}
{\cal Z} \odot {\cal Z}^\prime = q_0 q_0^\prime - \Vq \cdot \Vq^\prime
+ q_0 \Vq^\prime + q_0^\prime \Vq + \Vq \times \Vq^\prime,
\end{equation}
where $\odot$ denotes a quaternion multiplication.

A vector $\Vu$ can be rotated by the unitary transformation ${\cal Z}
\odot \Vu \odot {\cal Z}^{-1}$, where the multiplicative inverse of a
unit quaternion is ${\cal Z}^{-1} = q_0 - \Vq$.  Applying
Eq.~\ref{eq:quatprod} and treating $\Vu$ as a quaternion with zero
scalar component, the rotated vector $\Vu^\prime$ is given by
\begin{equation} \label{eq:quatrot}
    \Vu^\prime = (q_0^2-\Vq\cdot\Vq)\Vu + 2\Vq\Vq\cdot\Vu +
    2q_0\Vq\times\Vu,
\end{equation}
and remains a pure vector. The rotation can also be written in matrix
form, $u_i^\prime = d_{i\alpha} u_\alpha$, with the director vectors
that form the rotation matrix $d_{i\alpha}$ as given in
Eq.~\ref{eq:di}.

An infinitesimal change in the directors is given by a rotation
$\delta \Vphi$:
\begin{equation}\label{eq:dd}
\delta \Vd_i = \delta \Vphi \times \Vd_i,
\end{equation}
with $\VOmega = \Ds \Vphi$ and $\Vomega = \Dt \Vphi$. The combination
of the original rotation ${\cal Z}$ and an additional infinitesimal
rotation $\delta {\cal Z} = 1 + \delta \Vphi/2$ can be found by
applying the rotations sequentially,
\begin{equation} \label{eq:quatdquat}
\Vu^\prime + \delta \Vu = \delta {\cal Z} \odot {\cal Z} \odot \Vu
\odot {\cal Z}^{-1} \odot \delta {\cal Z}^{-1} = {\cal Z}^\prime \odot
\Vu \odot {\cal Z}^{\prime -1}.
\end{equation}
The quaternion ${\cal Z}^\prime$ is found by multiplying the two
quaternions,
\begin{equation}
{\cal Z}^\prime = \delta {\cal Z} \odot {\cal Z} = q_0 - \Vq \cdot
        \frac{\delta \Vphi}{2} + \Vq + q_0 \frac{\delta \Vphi}{2} -
        \Vq \times \frac{\delta \Vphi}{2}.
\end{equation}
Thus, the variation in the quaternion $\delta {\cal Z} = {\cal
Z}^\prime - {\cal Z}$ is linearly related to $\delta \Vphi$,
\begin{equation} \label{eq:dq}
        \left(
                \begin{array}{c}
                         \delta q_0 \\ \delta q_x \\ \delta q_y \\
                         \delta q_z
                \end{array}
        \right) = \frac{1}{2} \left(
                \begin{array}{ccc}
                        -q_x &  -q_y &  -q_z\\ ~~q_0 & ~~q_z &  -q_y\\
                        -q_z & ~~q_0 & ~~q_x\\ ~~q_y &  -q_x & ~~q_0
                \end{array}
        \right) \cdot \left(
                \begin{array}{c}
                        \delta\Vphi_x \\ \delta\Vphi_y \\
                        \delta\Vphi_z
                \end{array}
        \right) .
\end{equation}
The column vectors in Eq.~\ref{eq:dq} define a set of basis vectors in
the quaternion space, $e_{\alpha a}$, where $e_{\alpha a}$ is the
transpose of the matrix in Eq.~\ref{eq:dq}. These basis vectors are
orthogonal to $q_a$ and relate changes in quaternions to rotations
about the space-fixed axes,
\begin{equation} \label{eq:qp}
\delta \phi_\alpha = 2 e_{\alpha a} \delta q_a , ~~~ \delta q_a =
\frac{1}{2} e_{\alpha a} \delta \phi_\alpha .
\end{equation}
In this work we have used body-fixed rotations,
Eqs.~\ref{eq:q2phi}--\ref{eq:phi2q}, for which we need the basis
vectors $e_{ia}$ given in Eq.~\ref{eq:ei}; they are related to the
space fixed basis $e_{\alpha a}$ by the rotation matrix, $e_{i a} =
d_{i\alpha} e_{\alpha a}$. The vectors $e_{ia}$ or $e_{\alpha a}$,
together with $q_a$, form a complete basis in the quaternion space.

Finally, we obtain the derivatives of the basis vectors quoted in
Eqs.~\ref{eq:ddi}--\ref{eq:dei}. A variation in the basis vectors
$\Vd_i$ is related to an infinitesimal rotation, Eq.~\ref{eq:dd},
\begin{equation} \label{eq:vardi}
\delta d_{i \alpha} = \epsilon_{\alpha \beta \gamma} \delta \phi_\beta
d_{i \gamma} = \sum_{j,k=1}^3 \epsilon_{ijk} d_{j \alpha} \delta
\phi_k =  2 \sum_{j,k=1}^3 \epsilon_{ijk} d_{j \alpha} e_{kb} \delta
q_b.
\end{equation}
The variation in $\Vd_i$ can also be directly related to constrained
variations in quaternions,
\begin{equation}\label{eq:vardi2}
\delta d_{i \alpha} = \frac{\partial d_{i \alpha}}{\partial q_a}
\left( \delta_{ab} - q_a q_b \right) \delta q_b,
\end{equation}
where the projection operator $\left( \delta_{ab} - q_a q_b \right)$
is included to ensure that the normalization condition, $\delta q_a
q_a = 0$, is satisfied.  Equation~\ref{eq:ddi} can then be obtained by
making use of the result
\begin{equation}
q_a \frac{\partial d_{i \alpha}}{\partial q_a} = 2 d_{i \alpha}.
\end{equation}

The rotation matrix can be written as a product of $\Ve$ vectors,
$d_{i \alpha} = e_{i a}e_{\alpha a}$. A space fixed vector is first
rotated into the quaternion basis by $e_{\alpha a}/2$ and then rotated
from the quaternion basis to the body-fixed frame by $2e_{i a}$. A
variation in $d_{i \alpha}$ is then composed of two equal
contributions from variations in $e_{i a}$ and $e_{i \alpha}$,
\begin{equation} \label{eq:vardiei}
\delta d_{i \alpha} = \delta e_{i a} e_{\alpha a} + e_{i a} \delta
e_{\alpha a} = 2 \delta e_{i a} e_{\alpha a}.
\end{equation}
Substituting Eq.~\eqref{eq:vardi} for the variation in $d_{i \alpha}$,
and using the orthogonality of the $\Vd$ vectors,
\begin{equation} \label{eq:varei}
\delta e_{ia} e_{ja} =  \sum_{k=1}^3 \epsilon_{ijk} e_{ka}  \delta q_a.
\end{equation}
Multiplying both sides by $e_{j b}$ and summing over $j$,
\begin{equation} \label{eq:varei2}
 \left( \delta_{ab} - q_a q_b \right) \delta e_{ia} = \sum_{j,k=1}^3
 \epsilon_{ijk} e_{jb} e_{ka} \delta q_a.
\end{equation}
The variation in $e_{ia}$ can also be related to constrained variations in
$q_a$, \cf Eq.~\eqref{eq:vardi2}, using the relation $q_a \delta
e_{ia} =- e_{ia} \delta q_a$,
\begin{equation}\label{eq:varei3}
\delta e_{ia} = \frac{\partial e_{ia}}{\partial q_c} \left(
\delta_{bc} - q_b q_c \right) \delta q_b =  \sum_{j,k=1}^3
\epsilon_{ijk} e_{ja} e_{kb} \delta q_b - q_a e_{ib} \delta q_b.
\end{equation}
Equation~\ref{eq:dei} then follows from
\begin{equation}
q_b \frac{\partial e_{ia}}{\partial q_b} =  e_{ia}.
\end{equation}

\bibliographystyle{aip} \bibliography{filaments,md}

\end{document}